\documentclass[preprint, sort&compress]{elsarticle}

\usepackage{lineno,hyperref}
\modulolinenumbers[5]
\usepackage{graphicx}
\usepackage{dcolumn}
\usepackage{bm}
\usepackage{amsmath}
\usepackage{color}
\usepackage[toc,page]{appendix}

\usepackage{numcompress}
\newenvironment{rcases}
  {\left.\begin{aligned}}
  {\end{aligned}\right\rbrace}

\journal{Journal of Sound and Vibration}

\begin{document}

\begin{frontmatter}

\title{Formation of Local Resonance Band Gaps in Finite Acoustic Metamaterials: A Closed-form Transfer Function Model}


\author[mysecondaryaddress]{H. Al Ba'ba'a}

\author[mysecondaryaddress]{M. Nouh\corref{mycorrespondingauthor}}
\cortext[mycorrespondingauthor]{Corresponding author}
\ead{mnouh@buffalo.edu}

\author[mysecondaryaddress]{T. Singh}

\address[mysecondaryaddress]{Dept. of Mechanical \& Aerospace Engineering, University at Buffalo (SUNY), Buffalo, NY}

\begin{abstract}
The objective of this paper is to use transfer functions to comprehend the formation of band gaps in locally resonant acoustic metamaterials. Identifying a recursive approach for any number of serially arranged locally resonant mass in mass cells, a closed form expression for the transfer function is derived. Analysis of the end-to-end transfer function helps identify the fundamental mechanism for the band gap formation in a finite metamaterial. This mechanism includes (a) repeated complex conjugate zeros located at the natural frequency of the individual local resonators, (b) the presence of two poles which flank the band gap, and (c) the absence of poles in the band-gap. Analysis of the finite cell dynamics are compared to the Bloch-wave analysis of infinitely long metamaterials to confirm the theoretical limits of the band gap estimated by the transfer function modeling. The analysis also explains how the band gap evolves as the number of cells in the metamaterial chain increases and highlights how the response varies depending on the chosen sensing location along the length of the  metamaterial. The proposed transfer function approach to compute and evaluate band gaps in locally resonant structures provides a framework for the exploitation of control techniques to modify and tune band gaps in finite metamaterial realizations.
\end{abstract}


\end{frontmatter}


\section{Introduction}

\noindent 

Acoustic metamaterials (AMMs) are sub-wavelength structures that consist of chains of self-repeating unit cells which house internal elastic resonators. The hallmark feature of AMMs is their ability to realize band gaps, i.e. regions of blocked wave propagation, in low frequency regimes. Band gaps in AMMs primarily depend on the resonator properties and are, thus, size-independent and mechanically tunable \cite{liu_sonic}. Unique wave propagation behavior in AMMs have rendered them appealing for a wide range of damping and noise control applications. Over the past few decades, AMMs have been investigated in the context of discrete lumped mass systems \cite{huang2011study, Huang2010}, elastic bars \cite{wang2004, xiao2012longitudinal}, flexural beams \cite{yu2006a, yu2006b, sun2010, Nouh2014, Pai2014, hussein_AIP, xiao2013flexural, Baravelli2013, Zhu2014}, as well as 2D membranes and plates \cite{towards, pai3, wang2013, Nouh2015}. Given the dependence of the band structure of the AMM unit cell on resonator parameters, multiple efforts have also been placed on piezoelectric, or actively controlled, metamaterials \cite{Gonella2009, Celli2015, Chen2015, chen2014piezo, Nouh2016}.

Acoustic metamaterials are most commonly modeled using a Bloch-wave propagation model of the self-repeating unit cell with periodic boundary conditions \cite{Bloch1929, Hussein2014, Mead1970, Mead1971, Faulkner1985}. Wave propagation methods assume traveling wave propagation in an infinitely-long metamaterial comprised of the individual unit cells. The occurrence of band gaps in these infinite structures has been explained in light of gaps in the unit cell's dispersion curve (band diagram) and/or the negative effective mass density concept \cite{huang2009negative, Pai2010}. Discrepancies in the response of actual metamaterials motivated several efforts to understand band gap realizations in finite structures and the effect of imposed boundary conditions \cite{gupta1970natural,nielsen2015periodicity,hvatov2015free}. Significant among those is the investigation of the relationship between the borders of Bragg-effect band gaps in phononic (periodic) structures and the corresponding eigenfrequencies, explained using the phase-closure principle \cite{nielsen2015periodicity,hvatov2015free}. Modal analysis has also been utilized to develop a mathematical formulation to estimate locally resonant band gaps and provide design guidelines and insights into the choice of resonators and their optimal locations \cite{sugino2016mechanism}. To this date, however, a mathematical framework that explains and quantifies the evolution of local resonance band gaps in finite AMMs remains lacking. 

\begin{figure}[h]
\centering
\includegraphics[width=0.75\textwidth]{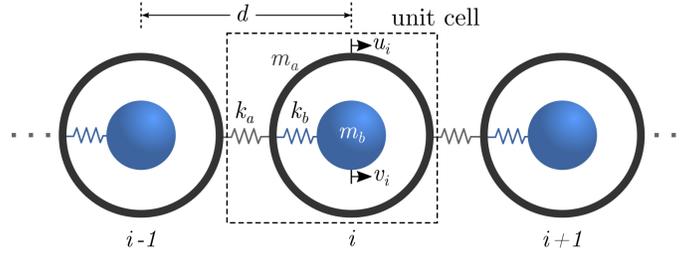}
\caption{\label{fig:acoustic_mm_gray} A lumped mass-in-mass locally resonant acoustic metamaterial}
\end{figure}

In this effort, we focus on AMMs where the number of cells, as well excitation and response locations, are specified. We derive a generalized dynamic model to evaluate the input-to-output transfer function associated with such locally resonant structures, and explain the formation mechanism of the band gap in light of their frequency response and pole-zero (PZ) distributions. To facilitate the discussion and advance a closed-form solution, the analysis is carried out on a one-dimensional mass-in-mass type metamaterial. The AMM consists of a chain of spring-mass unit cells shown in Figure \ref{fig:acoustic_mm_gray}. In the presented analysis, damping elements are excluded from both the base and the local structure for two important reasons: (1) to neutralize the effect of dissipation on the band gaps, an effect that has been recently investigated in a number of efforts \cite{hussein_damp1, hussein_damp2, huang_damp3, andreassen_damp4}, and (2) to ensure that any damping captured in the numerically computed poles or zeros in the lengthy expressions of the developed dynamic model of the finite AMM are merely a result of minor errors or computational precision, as will be highlighted later in the discussion. The limiting case of the presented approach as the length of the AMM chain approaches infinity matches the traditional Bloch-wave analysis and bridges the gap between the two approaches.

Finally, the discussion is extended to explain and differentiate between the effects of sensing location (i.e., location where displacement is measured) and the effect of the number of cells on the bandwidth and degree of attenuation obtained from the local resonators. Analysis of AMMs from a dynamic systems perspective provides a physical insight into the formation of these band gaps over a specific range of frequencies, and provides a clear distinction between the operation concepts of AMMs and tuned dynamic absorbers from a vibrations standpoint. Furthermore, explaining the behavior in terms of frequency domain tools and PZ maps sets a future framework for implementing control techniques. Finally, the investigation of finite metamaterial structures is naturally of interest since the results directly impact the fabrication of realistic and physically realizable, rather than purely theoretical, AMMs.

\section{Dynamics of 1-D Acoustic Metamaterials}
\subsection{\label{sec:Wave_Disp} Wave Dispersion Analysis}
The simplest example of an AMM is a periodic series of spring-mass systems hosting internal spring-mass resonators, as shown in Figure \ref{fig:acoustic_mm_gray}. For an AMM with cell spacing $d$, identical outer masses $m_a$ and inner masses $m_b$ connected via springs $k_a$ and $k_b$, the governing motion equations for the $i^{th}$ unit can be derived as:

\begin{equation}
\begin{bmatrix}
m_a & 0 \\ 0 & m_b
\end{bmatrix}
\begin{Bmatrix}
\ddot{u}_i \\ \ddot{v}_i
\end{Bmatrix}
+
\begin{bmatrix}
2k_a+k_b & -k_b \\ -k_b & k_b
\end{bmatrix}
\begin{Bmatrix}
u_i\\v_i
\end{Bmatrix}
+
\begin{bmatrix}
-k_a & -k_a \\ 0 & 0
\end{bmatrix}
\begin{Bmatrix}
u_{i-1} \\ u_{i+1}
\end{Bmatrix}
=
\begin{Bmatrix}
0\\0
\end{Bmatrix}
\label{EOMs}
\end{equation}

\begin{figure}[h]
\centering
\includegraphics[width=0.9\textwidth]{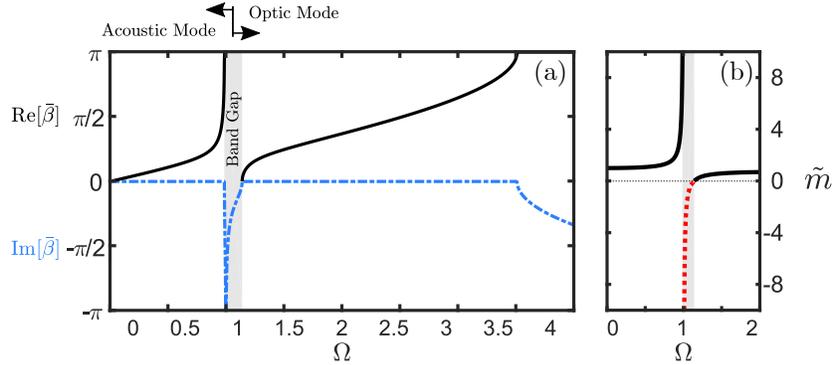}
\caption{\label{fig:Disp_NEM}(a) Band structure of the AMM unit cell obtained from its dispersion relations and (b) the corresponding normalized effective mass $\tilde{m} = \frac{m_{e}}{(m_a+m_b)}$. The shaded region highlights the band gap frequency span in (a) which matches the negative effective mass region in (b)}
\end{figure}

\vskip 1em
\noindent where $u_i$ and $v_i$ represent the displacements of $m_a$ and $m_b$ of the $i^{th}$ cell, respectively. By applying the harmonic wave solution to the above motion equations, the dispersion relation can be derived as \cite{Huang2010}
\begin{equation}
A_1 \omega^4 + A_2 \omega^2 + A_3 = 0
\label{disp}
\end{equation}
\noindent with
\begin{center}
$A_1=m_a m_b$, \hspace{0.09cm}
$A_2=-[(m_a+m_b)k_b+2m_bk_a(1-\cos \bar{\beta})]$, \hspace{0.09cm}
$A_3=2k_a k_b (1-\cos \bar{\beta})$\\
\end{center}

\noindent where $\omega$ is the angular frequency and $\bar{\beta}$ is the normalized wavenumber given by $2 \pi d/\lambda$ where $\lambda$ is the wavelength, or the spatial period of the propagating wave. Equation (\ref{disp}) can be normalized in terms of the mass ratio $m_r=m_b/m_a$, the stiffness ratio $k_r=k_b/k_a$, and a non-dimensional frequency $\Omega=\omega/\omega_b$ to give

\begin{equation}
\Omega^4 - [(1+m_r)+2\Gamma(1-\cos \bar{\beta})]\Omega^2 + 2\Gamma(1-\cos \bar{\beta}) = 0
\label{disp_2}
\end{equation}

\noindent where $\omega_b=\sqrt{k_b/m_b}$ is the natural frequency of the local resonator and $\Gamma = \frac{m_r}{k_r}$. Eq. (\ref{disp_2}) can be used to compute the dispersion curves (band structure) $\Omega(\bar{\beta})$ of the AMM. Figure \ref{fig:Disp_NEM}(a) shows these curves for an AMM with \(m_a=1\) kg, \(m_b=0.3 m_a\), \(k_a=4.8\) GN/m and \(k_b=0.1 k_a\). The shaded region represents the frequency range of the local resonance band gap which also spans the negative effective mass $m_{e}$ region of an equivalent homogeneous material, as shown in Figure \ref{fig:Disp_NEM}(b), and given by \cite{huang2009negative}

\begin{equation}
m_e=m_a+\frac{m_b \omega_b^2}{\omega_b^2-\omega^2}.
\label{effective_mass}
\end{equation}

\noindent The band gap splits the dispersion curve of the metamaterial into acoustic and optic branches (where $\bar{\beta}$ is purely real) representing in-phase and out-of-phase propagating wave modes in the outer mass and the internal resonator. The theoretical bounds of the band gap can be obtained directly from Eq. (\ref{disp_2}). By setting $\bar{\beta} = \pi$, the solution of the dispersion relation yields the lower bound $\Omega_l$ which is, for this case, equal to 0.9867. Generally, $\Omega_l$ will vary with both the stiffness and the mass ratios of the individual cell and is given by

\begin{equation}
\Omega_l = \frac{1}{\sqrt{2}}\sqrt{(1+m_r+4\Gamma)-\sqrt{(1+m_r+4\Gamma)^2-16 \Gamma}}
\label{eq:Omegal}
\end{equation}

\noindent On the other hand, the upper bound of the band gap $\Omega_u$ can be obtained by setting $\bar{\beta} = 0$ to obtain a non-zero solution of
\begin{equation}
\Omega_u = \sqrt{1+m_r}
\label{eq:Omegau}
\end{equation}

\noindent which, unlike the lower bound, solely depends on the mass ratio $m_r$. 

\begin{figure}[h]
\centering
\includegraphics[width=0.7\textwidth]{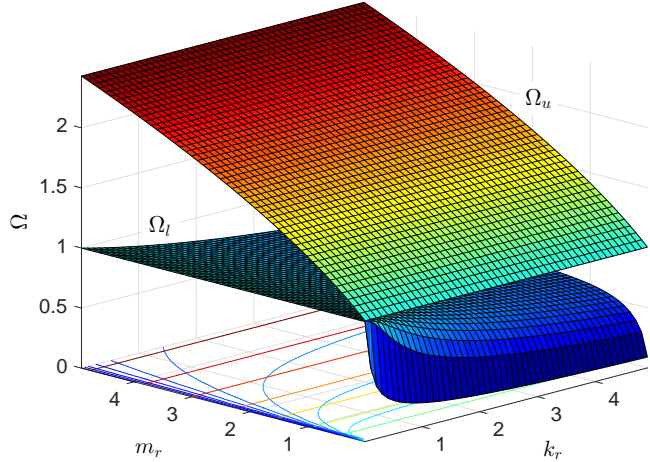}
\caption{\label{fig:kr_mr_limits} Variation of the local resonance band gap bounds, $\Omega_l$ and $\Omega_u$, with $m_r$ and $k_r$}
\end{figure}

Figure \ref{fig:kr_mr_limits} graphically depicts the effect of both the mass and stiffness ratios on the local resonance band gap bounds, which will be relevant to the discussion in Sec. \ref{Numerical}. Important to note here is that the AMM band gap does not necessarily start at the natural frequency of the local resonator (i.e. at $\Omega_l=1$), which is a common misconception. It can be also seen that as $k_r$ approaches zero, the solution of $\Omega_l$ in Eq. (\ref{eq:Omegal}) approaches 1. However, a larger $k_r$ shifts the lower end of the band gap down, effectively shortening the acoustic mode. In which case, the AMM band gap (and attenuation of incident waves) can start at a frequency lower than the local resonance. Although existent, the effect of $m_r$ on $\Omega_l$ is however minimal for $m_r$ values larger than 1. On the other hand, the upper bound of the band gap $\Omega_u$ is solely influenced by $m_r$ which also shapes the final structure of the optic branch. Finally, it is worth noting that an exact $\Omega_l=1$ value is not achievable from Eq. (\ref{eq:Omegal}) for any combination of $m_r$ and $k_r$. However, $\Omega_l=0$ may be obtained as a result of a zero $\Gamma$ and a non-zero $m_r$, (i.e. an infinitely stiff local spring), which is a purely theoretical case.

\subsection{Structural dynamics of a finite AMM}
\label{Sec:EOM_AMM}
Since the AMM in Figure \ref{fig:acoustic_mm_gray} is described by a series of discrete mechanical elements, the profile of the propagating waves is captured by the discretized displacement field given by the vectors \textbf{u} and \textbf{v} along the length $L$ of the AMM, where 

\begin{equation}
\mathbf{u}=
\begin{Bmatrix}
u_1 & u_2 & & ... && u_i && ... && u_n
\end{Bmatrix}
^T
\label{xs}
\end{equation}

and

\begin{equation}
\mathbf{v}=
\begin{Bmatrix}
v_1 & v_2 & & ... && v_i && ... && v_n
\end{Bmatrix}
^T
\label{vs}
\end{equation}

\noindent For an AMM of \(n\) cells, the equations of motion of the end cells slightly differ from the middle ones. For any $i^{th}$ cell where \(2 \leq i \leq n-1\), the motion is given by 

\begin{equation}
\begin{gathered}
 m_a \ddot{u}_i+(2 k_a+k_b) u_i -k_a u_{i-1} -k_a u_{i+1} - k_b v_i = 0
\\
 m_b \ddot{v}_i + k_b v_i - k_b u_i = 0
\end{gathered}
\label{middle_cells}
\end{equation}

\noindent while the first (\(i=1\)) and last (\(i=n\)) cells are described by

\begin{equation}
\begin{gathered}
 m_a \ddot{u}_1+(k_a+k_b) u_1 -k_a u_{2} - k_b v_1 = F
\\
 m_b \ddot{v}_1 + k_b v_1 - k_b u_1 = 0
\end{gathered}
\label{first_cell}
\end{equation}

\noindent and

\begin{equation}
\begin{gathered}
 m_a \ddot{u}_n+(k_a+k_b) u_n -k_a u_{n-1} - k_b v_n = 0
\\
 m_b \ddot{v}_n + k_b v_n - k_b u_n = 0
\end{gathered}
\label{last_cell}
\end{equation}

\noindent where $F$ represents the forcing acting on the outer mass of the first unit cell. As a result, the steady-state oscillations of a finite AMM subject to the external sinusoidal excitation \textbf{F} is given by 

\begin{equation}
\begin{Bmatrix}
\mathbf{u} \\ \mathbf{v}
\end{Bmatrix}
=
\Bigg(
- \omega^2
\begin{bmatrix}
\mathbf{M_u} & \mathbf{0} \\ \mathbf{0} & \mathbf{M_v}
\end{bmatrix}
+
\begin{bmatrix}
\mathbf{K_u} & \mathbf{-K_v} \\ \mathbf{-K_v} & \mathbf{K_v}
\end{bmatrix}
\Bigg)
^{-1}
\begin{Bmatrix}
\mathbf{F}\\ \mathbf{0}
\end{Bmatrix}
\label{EOM_global}
\end{equation}

\noindent where the sub-components of the mass and stiffness matrices can be obtained from

\begin{align}
\underset{n\times n}{\mathrm{\mathbf{M_u}}}=
\begin{bmatrix}
m_a  &  &  & &      &   \\
& m_a & & & &\\
 &  & \ddots &  &  &    \\
& & & & & m_a\\
\end{bmatrix} && 
\underset{n\times n}{\mathrm{\mathbf{M_v}}}=
\begin{bmatrix}
m_b  &  &  & &      &   \\
& m_b & & & &\\
 &  & \ddots &  &  &    \\
& & & & & m_b\\
\end{bmatrix}
\label{matrices1}
\end{align}

\begin{align}
\underset{n\times n}{\mathrm{\mathbf{K_v}}}=
\begin{bmatrix}
k_b  &  &  & &      &   \\
& k_b & & & &\\
 &  & \ddots &  &  &    \\
& & & & & k_b\\
\end{bmatrix} 
\label{matrices2}
\end{align}

\begin{align}
\underset{n\times n}{\mathrm{\mathbf{K_u}}}=
\begin{bmatrix}
k_a+k_b  & -k_a  &  & &      &   \\
-k_a & 2k_a+k_b & -k_a & & \\
 & \ddots  & \ddots &  \ddots &  &    \\
 &  &  -k_a & 2k_a+k_b &  -k_a&    \\
& & & -k_a & k_a+k_b\\
\end{bmatrix} 
\label{matrices3}
\end{align}

\noindent while \textbf{F} is equal to

\begin{equation}
\underset{n\times 1}{\mathrm{\mathbf{F}}}=
\begin{Bmatrix}
F & 0 & \dots & 0
\end{Bmatrix}
^T
\label{vs}
\end{equation}

\subsection{Analytical formula for the AMM's natural frequencies}
The free vibration of an $n$-cell AMM is given by

\begin{equation}
\begin{bmatrix}
\mathbf{M_u} & \mathbf{0} \\ \mathbf{0} & \mathbf{M_v}
\end{bmatrix}
\begin{Bmatrix}
\mathbf{\ddot{u}} \\ \mathbf{\ddot{v}}
\end{Bmatrix}
+
\begin{bmatrix}
\mathbf{K_u} & \mathbf{-K_v} \\ \mathbf{-K_v} & \mathbf{K_v}
\end{bmatrix}
\begin{Bmatrix}
\mathbf{u} \\ \mathbf{v}
\end{Bmatrix}
=
\begin{Bmatrix}
\mathbf{0} \\ \mathbf{0}
\end{Bmatrix}
\label{EOM_freevib}
\end{equation}

Assuming harmonic motion, and normalizing the excitation frequency with $\omega_b$, we obtain

\begin{equation}
\begin{bmatrix}
\mathbf{\Omega}-\Gamma \mathbf{\Psi} - m_r\mathbf{I} & m_r\mathbf{I} \\ 
\mathbf{I} & \mathbf{\Omega}-\mathbf{I}
\end{bmatrix}
\begin{Bmatrix}
\mathbf{u} \\ \mathbf{v}
\end{Bmatrix}
=
\begin{Bmatrix}
\mathbf{0} \\ \mathbf{0}
\end{Bmatrix}
\label{EOM_free_normalized}
\end{equation}

\noindent where $\mathbf{I}$ is the identity matrix and $\mathbf{\Omega}$ and $\mathbf{\Psi}$ are given by

\begin{align}
\underset{n\times n}{\mathrm{\mathbf{\Omega}}}=
\begin{bmatrix}
\Omega^2  &  &  & &      &   &\\
& \Omega^2 & & & & &\\
 &  &  &  &  &   & \\
  &  &  & \ddots &  &  &  \\
   &  &  &  &  &   & \\
   &  &  &  &  & &   \\
& & & & & & \Omega^2\\
\end{bmatrix} 
\label{eq:Omega}
\end{align}

\begin{align}
\underset{n\times n}{\mathrm{\mathbf{\Psi}}}=
\begin{bmatrix}
1  & -1 &  & &      &   &\\
-1 & 2 & -1 & & & &\\
 &  &  &  &  &   & \\
  &  & \ddots & \ddots & \ddots &  &  \\
   &  &  &  &  &   & \\
   &  &  &  & -1  & 2 & -1  \\
& & & & & -1 & 1\\
\end{bmatrix} 
\label{eq:Psi}
\end{align}

\noindent It is seen from Eq. (\ref{EOM_free_normalized}) that $\mathbf{u}+\mathbf{(\Omega-I)}\mathbf{v} = \mathbf{0}$, which allows us to reduce the left-hand matrix to be in terms of $\mathbf{u}$ only

\begin{equation}
\begin{bmatrix}
\mathbf{\Omega^2}-\big(\Gamma \mathbf{\Psi} + (m_r+1) \mathbf{I}\big) \mathbf{\Omega} + \Gamma \mathbf{\Psi}
\end{bmatrix}
\begin{Bmatrix}
\mathbf{u}
\end{Bmatrix}
=
\begin{Bmatrix}
\mathbf{0}
\end{Bmatrix}
\label{EOM_free_norm_reduced}
\end{equation}

\noindent where the dimensions of the new matrix are $n \times n$. The matrix in Eq. (\ref{EOM_free_norm_reduced}) is a symmetric tridiagonal matrix with the same diagonal elements $b$, except for the first and last terms, which may be written in the following general form

\begin{equation}
\begin{bmatrix}
-\eta+b & a & & & & &\\
a & b & a & & & & \\
& & & & & &\\
& & \ddots &\ddots & \ddots &  & \\
& & & & & &\\
& & & & a & b & a &\\
& & & & & a & -\xi+b\\
\end{bmatrix}
\end{equation}

\noindent where $a = \Gamma (\Omega^2-1)$ and $b = \Omega^4-(1+m_r+2\Gamma)\Omega^2+2\Gamma$. It has been reported that for values of $\xi$ and $\eta$ equal to $-a$, which is the case here, the eigenvalues for such a matrix takes the following form \cite{yueh2005eigenvalues}

\begin{equation}
    \lambda_k = b +2a\cos \theta_k
    \label{eq:General_EVP}
\end{equation}

\noindent where $k = 1,2, \dots, n$ and $\theta_k=\frac{k-1}{n}\pi$. Once the eigenvalues are established, the left-hand matrix in Eq. (\ref{EOM_free_norm_reduced}) can be decomposed into $\mathbf{Q}\mathbf{\Lambda_k}\mathbf{Q^{T}}$ since the eigenvectors of a symmetric matrix are orthogonal. $\mathbf{\Lambda_k}$ is the eigenvalue matrix and $\mathbf{Q}$ and $\mathbf{Q^{T}}$ are the corresponding left and right eigenvector matrices, respectively. The decomposed matrix facilitates the calculation of the determinant of the original matrix (Eq. (\ref{EOM_free_norm_reduced})). Since $|\mathbf{Q} \mathbf{Q^{T}}|=1$, $|\mathbf{\Lambda_k}|$ has to be zero and the natural frequencies of an $n$-cell AMM are found as

\begin{equation}
    \Omega^4_k - \big(1+m_r+2\Gamma (1-\cos \theta_k)\big)\Omega^2_k +2\Gamma (1-\cos \theta_k) = 0
    \label{eq:analytical_poles}
\end{equation}

\noindent Each value of $k$ gives two real positive solutions for $\Omega^2_k$, resulting in $2n$ values of the natural frequencies. For convenience, we define a new index $q$ to denote the $2n$ values of natural frequencies $\Omega_q$, where $q = 1,2, \dots ,2n$ and $\Omega_1<\Omega_2< \dots < \Omega_{2n}$. Each $k$ gives two natural frequencies at $q=k$ and $q=k+n$, resulting in $\Omega_k$ and $\Omega_{k+n}$, which correspond to the acoustic and optic modes respectively. The eigenvectors $\mathbf{u}^{(q)}$ are given by \cite{yueh2005eigenvalues}

\begin{equation}
    u^{(q)}_i = u^{(k)}_i = u^{(k+n)}_i =\rho^{i-1}
    \label{eq:eigen_vec_2_free_1}
\end{equation}

\noindent for $k=1$, and

\begin{equation}
    u^{(q)}_i = u^{(k)}_i = u^{(k+n)}_i = \rho^{i-1} \cos \frac{(2k-1)(2i-1)\pi}{2n}
    \label{eq:eigen_vec_2_free_n}
\end{equation}

\noindent for $k=2,3, \dots, n$, and $\rho = 1$. Next, the eigenvectors of the local resonators $\mathbf{v}^{(q)}$ can be found as

\begin{equation}
    \mathbf{v}^{(q)}= (\mathbf{I}-\mathbf{\Omega}_q)^{-1} \mathbf{u}^{(q)}
    \label{eq:eigen_vec_v}
\end{equation}

\noindent and as a result the complete eigenvector of the AMM $\mathbf{Q}^{(q)}$ is given by

\begin{equation}
    \mathbf{Q}^{(q)}=\{\mathbf{u}^{(q)} \, \mathbf{v}^{(q)}\}^T
    \label{eq:eigen_vec_Q}
\end{equation}

\noindent Several observations can be made from the above set of equations. First, the eigenvectors (oscillation modes) of the outer and local masses, $\mathbf{u}^{(q)}$ and $\mathbf{v}^{(q)}$, are related by the scalar $(\mathbf{I}-\mathbf{\Omega}_q)^{-1}$ which depends on the eigenfrequency $\Omega_q$. It is evident that for $\Omega_q<1$, $(1-\Omega_q)$ becomes positive and consequently the outer and inner oscillate in-phase (confirming the commonly known acoustic mode in Figure \ref{fig:Disp_NEM}). The opposite happens for $\Omega_q>1$ leading to out-of-phase motion, or the optic mode. Second, Eqs. (\ref{eq:eigen_vec_2_free_1}) and (\ref{eq:eigen_vec_2_free_n}) reveal that the two eigenfrequencies $\Omega_k$ and $\Omega_{k+n}$ have the same outer mass eigenvector (i.e. $\mathbf{u}^{(k)}=\mathbf{u}^{(k+n)}$). Nonetheless, the complete eigenvectors $\mathbf{Q}^{(q)}$ are still unique since $\mathbf{v}^{(q)}$ is different for each value of $\Omega_q$ as depcited in Eq. (\ref{eq:eigen_vec_v}). 

\begin{figure}[h]
\centering
\includegraphics[width=\textwidth]{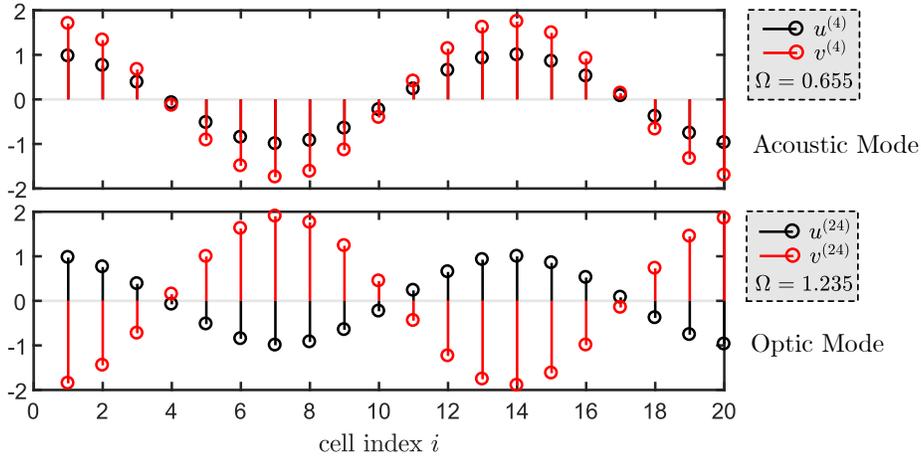}
\caption{Mode shapes of a 20-cell AMM for $k=4$ }
\label{fig:Modes}
\end{figure}

Figure \ref{fig:Modes} shows the eigenvectors for $k=4$ of an AMM with 20 cells (i.e. $q=4$ and $q=24$). As predicted, $u^{(4)}$ and $v^{(4)}$ oscillate in-phase, while $u^{(24)}$ and $v^{(24)}$ are out-of-phase with $u^{(24)}$ being identical to $u^{(4)}$. Finally, it is worth noting that when $k=1$ (which corresponds to the two natural frequencies $\Omega_1 = 0$ and $\Omega_{1+n} = \sqrt{1+m_r}$), Eq. (\ref{eq:eigen_vec_2_free_1}) yields identical eigenvectors of ones for  $\mathbf{u}^{(1)}$ and $\mathbf{u}^{(1+n)}$. In the $\Omega_1 = 0$ case, this understandably represents a rigid body mode for the unconstrained AMM. For $\Omega_{1+n} = \sqrt{1+m_r}$, however, all the outer masses $m_a$ of the AMM oscillate as a single body. In other words, the relative motion between the masses is absent and the connecting springs $k_a$ do not deform. Interestingly, the same happens in the inner masses or the local resonators of the AMM. In this scenario, the motions $\mathbf{u}^{(1+n)}$ and $\mathbf{v}^{(1+n)}$ are out-of-phase such that $\mathbf{v}^{(1+n)}=\frac{-1}{m_r}\mathbf{u}^{(1+n)}$.

While this section derived an analytical solution for the eigenvalues and corresponding eigenvectors for a finite acoustic metamaterial, the following section presents a transfer function formulation to represent the input-output characteristics of a finite AMM with the objective of using frequency response characteristics to interpret the creation mechanisms of local resonance band gaps.

\section{Transfer Functions and Pole-Zero Distributions {\label{TFs}}}

\subsection{Transfer function of a 2-cell AMM}

\begin{figure}[h]
\centering
\includegraphics[width=\textwidth]{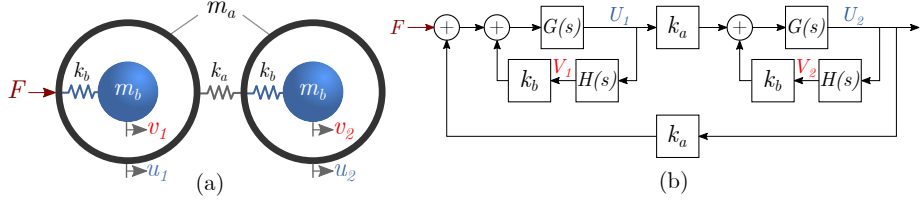}
\caption{(a) Schematic and (b) Block diagram of a 2-cell locally resonant AMM}
\label{fig:Cell}
\end{figure}

Starting with a 2-cell AMM of the same unit cell configuration (Figure \ref{fig:Cell}(a)), we derive the transfer function relating the displacement of the second outer mass $u_2$ to an input force $F$ applied to the first. The 2-cell AMM constitutes a 4 degree-of-freedom (DOF) system, with the DOFs being the outer and inner mass displacements of the 2 cells. Following Eqs. (\ref{first_cell}) and (\ref{last_cell}), we can rewrite the motion equations in the Laplace domain via

\begin{equation}\label{eq:TF1}
\textit {$U_1 = {G(s)}(k_a U_2+k_b V_1+F(s))$}
\end{equation}

\begin{equation}\label{eq:TF2}
\textit {$U_2 ={G(s)}(k_a U_1+k_b V_2)$}
\end{equation}

\begin{equation}\label{eq:TF3}
\textit {$\frac{V_2}{U_2} = \frac{V_1}{U_1} = H(s)$}
\end{equation}

\noindent where the upper case $U_i(s)$ and $V_i(s)$ represent the displacements $u_i(t)$ and $v_i(t)$ in the Laplace domain, written as $U_i$ and $V_i$ for brevity, and the transfer functions $G(s)$ and $H(s)$ are given by $\frac{1}{m_a s^2+(k_a+k_b)}$ and $\frac{k_b}{m_b s^2+k_b}$, respectively. The dynamics of the 2-cell AMM can be graphically represented using the block diagram shown in Figure \ref{fig:Cell}(b). The end-to-end transfer function $U_2/F$ can be obtained from Eqs. (\ref{eq:TF1}) through (\ref{eq:TF3}) or by reducing the shown block diagram. We will use the block diagram reduction approach as it reveals some trends and consistent patterns in the dynamics of this class of locally resonant metamaterials that will be later used to generalize this framework for an AMM with any number of cells $n$.

\begin{figure}[h]
\includegraphics[width=\textwidth]{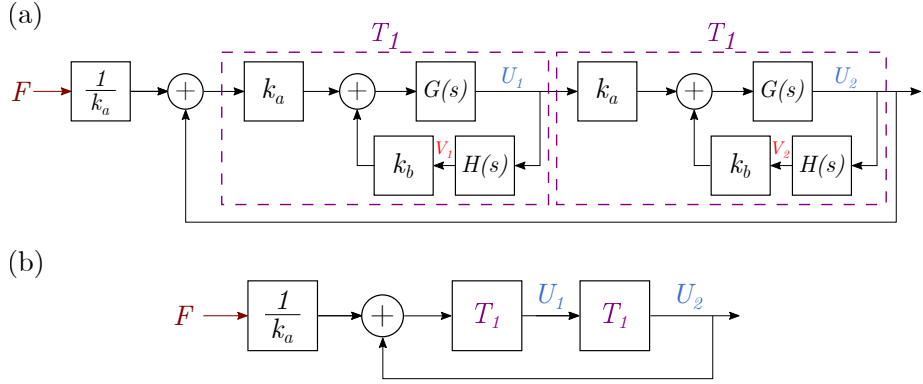}
\centering
\caption{Block diagram of the 2-cell AMM reduced from Figure \ref{fig:Cell}(b)}
\label{fig:BD2Cells_2}
\end{figure}

\noindent Moving the feedback gain $k_a$ to the forward path, and making the necessary adjustments, reduces the system to the diagram in Figure \ref{fig:BD2Cells_2}(a), where the dashed boxes represent similar structures that can be replaced with the single transfer function $T_1$ as shown in Figure \ref{fig:BD2Cells_2}(b).

\noindent $T_1$ simplifies Eqs. (\ref{eq:TF1}) and (\ref{eq:TF2}) to the following form

\begin{equation}\label{eq:T1}
\textit {$U_1 = T_1(s)(U_2+F(s)/k_a)$}
\end{equation}
\begin{equation}\label{eq:T1}
\textit {$U_2 = T_1(s)U_1$}
\end{equation}

\noindent and is given by

\begin{equation}\label{eq:T1}
\textit {$T_1 = \frac{\alpha_1 s^2 + \alpha_2}{s^4+\alpha_3 s^2 + \alpha_2}$}
\end{equation}

\noindent where $\alpha_1 = \omega_a^2= \frac{k_a}{m_a}$, $\alpha_2 = \omega_a^2\omega_b^2$, and $\alpha_3 = \omega_a^2+\omega_b^2+k_b/m_a$. Reducing Figure \ref{fig:BD2Cells_2}(b) to a single block gives the transfer function $U_2/F$ 

\begin{equation}\label{eq:U2_F}
\textit {$\frac{U_2}{F} = \frac{(\alpha_1 s^2 + \alpha_2)^2}{k_as^2\big(s^2+(\alpha_3-\alpha_1)\big)\big(s^4+(\alpha_3+\alpha_1)s^2 + 2\alpha_2\big)}$}
\end{equation}

\begin{figure}[h]
\includegraphics[width=0.9\textwidth]{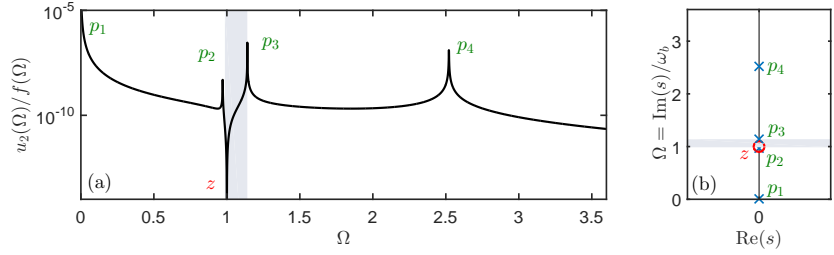}
\centering
\caption{(a) Frequency response of the outer displacement of the second cell $u_2$ in a 2-cell AMM as a ratio of the input force $f$. (b) Corresponding pole-zero map of the $U_2/F$ transfer function shown in Eq. (\ref{eq:U2_F}). $p$ and $z$ denote the locations of the poles and zeros, respectively. The shaded region in both plots represents the band gap span of an infinite AMM as predicted by the unit cell dispersion relations in Figure \ref{fig:Disp_NEM}}
\label{fig:2cell_bode_pz}
\end{figure}

The dynamics of the 2-cell AMM captured by Eq. (\ref{eq:U2_F}) is graphically represented in Figure \ref{fig:2cell_bode_pz}, for the same values of $m_a$, $m_b$, $k_a$, and $k_b$ used earlier to plot the dispersion curves in Figure \ref{fig:Disp_NEM}. As depicted in the transfer function, the AMM has two repeated poles at $s=0$ and at $s=\pm j\omega_b\sqrt{1+m_r}$ ($p_1$ and $p_3$ in Figure \ref{fig:2cell_bode_pz}). The poles at the origin are representative of the AMM's rigid body modes since the system considered is unconstrained (i.e. free-free). The poles at $\pm j\omega_b\sqrt{1+m_r}$ indicate the presence of a resonant frequency of the 4-DOF AMM right at the upper bound of the band gap, as given earlier by Eq. (\ref{eq:Omegau}). The significance of the remaining poles will later become clear when the general case of an AMM with a $n$ number of cells is discussed. On the other hand, the 2-cell AMM has repeated zeros at $s=\pm j\omega_b$ which is the stand-alone natural frequency of the local resonator. In this case, the 2-cell AMM acts as a vibration absorber with a tuned anti-resonance at $\omega_b$ where the response zeros out as shown by the frequency response in Figure \ref{fig:2cell_bode_pz}(a). The formation of a band gap that spans a frequency range rather than a single frequency is not evident at this point.

\subsection{General formulation for an $n$-cell AMM}
\begin{figure}[h]
\includegraphics[width=\textwidth]{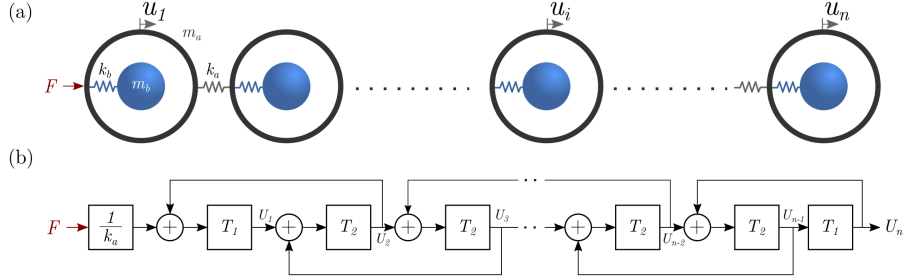}
\centering
\caption{(a) Schematic of a finite realization of AMM with $n$ number of cells and (b) its corresponding block diagram}
\label{fig:acoustic_mm_finite}
\end{figure}

Using a similar approach, and by utilizing the repeating patterns in the block structure of the AMM, a closed-form expression for the transfer function of a general lumped AMM with any given number of cells $n$ can be derived. The obtained transfer function can specifically describe the displacement of any $i^{th}$ cell in the AMM to a given forcing input. In the next steps, the exciting force $F$ is still applied to the outer mass of the first AMM cell. Owing to the periodic nature of the AMM, the block diagram of $n$ cells is now presented by a series of nested unity feedback loops with two forward path transfer functions $T_1$ and $T_2$, as shown in Figure \ref{fig:acoustic_mm_finite}. $T_2$ is given by

\begin{equation}\label{eq:T2}
\textit {$T_2 = \frac{\alpha_1 s^2 + \alpha_2}{s^4+(\alpha_1+\alpha_3) s^2 + 2 \alpha_2}$}
\end{equation}

\begin{figure}[h]
\includegraphics[width=\textwidth]{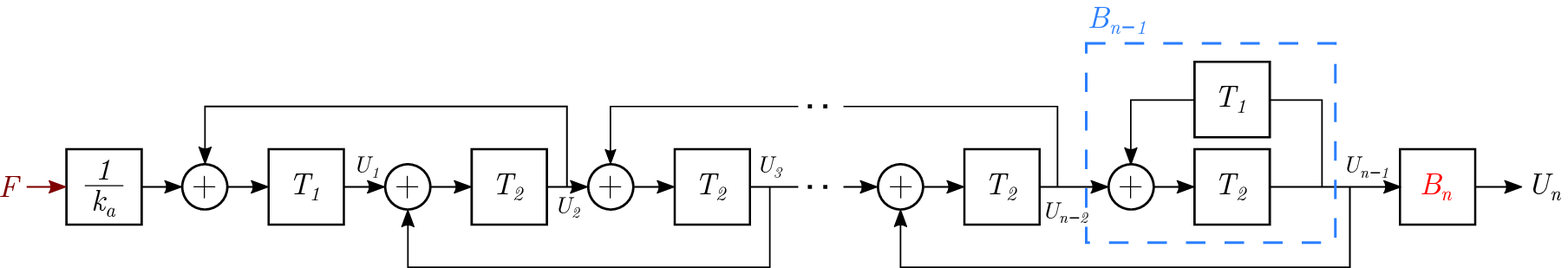}
\centering
\caption{Block diagram of an $n$-cell AMM reduced from Figure \ref{fig:acoustic_mm_finite}(b)}
\label{fig:Cont_frc}
\end{figure}

\noindent One way to obtain an expression for the end-to-end transfer function $U_n/F$ is to move the feedback branch from $U_n$ to the left of last $T_1$ block. A reduced feedback loop $B_{n-1}$, highlighted in Figure \ref{fig:Cont_frc}, can be generated as a result. Note that the last block in Figure \ref{fig:Cont_frc} is $B_n = T_1$. Moving backward and repeating this procedure yields another reduced block $B_{n-2}$ that is a function of $B_{n-1}$, and so on. This process can be repeated $n$ times going back all the way to the force input.

\begin{figure}[h]
\includegraphics[width=\textwidth]{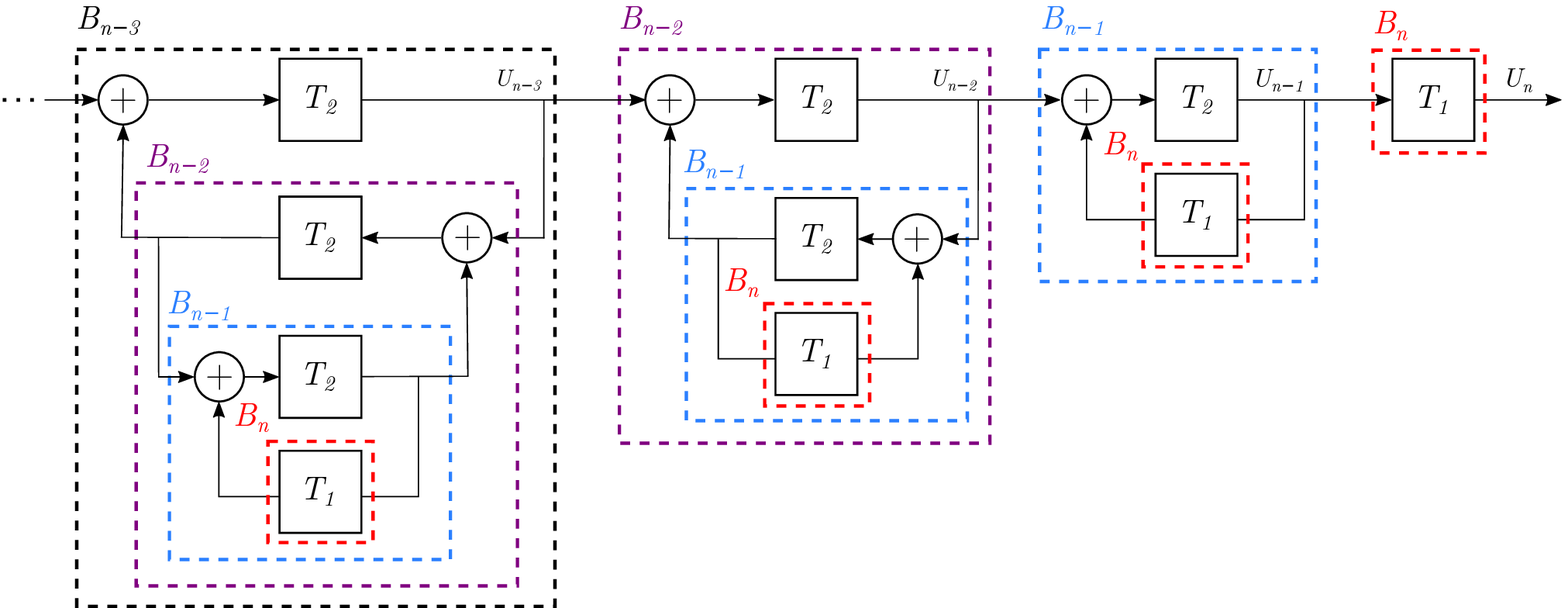}
\centering
\caption{A continuous fraction model for an $n$-cell AMM}
\label{fig:Cont_frc_2}
\end{figure}

\begin{figure}[h]
\includegraphics[width=\textwidth]{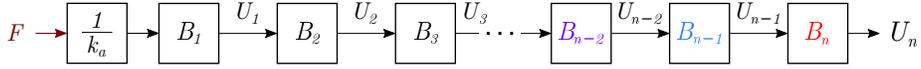}
\centering
\caption{Block diagram of an $n$-cell AMM reduced from Figure \ref{fig:Cont_frc_2}}
\label{fig:BD_R}
\end{figure}

This repetitive process of computing the equivalent transfer function of the closed loop is found to be analogous to the continuous fraction technique \cite{hensley_contfrac}, which can be used to obtain a general formula for the block $B_{n-(j+1)}$, as shown in Figure \ref{fig:Cont_frc_2}. The final form is shown in Figure \ref{fig:BD_R}.

\newpage \noindent Figure \ref{fig:Cont_frc_2} can be mathematically expressed as

\begin{equation*}
\frac{U_{n-(j+1)}}{U_{n-(j+2)}} = B_{n-(j+1)}=
\end{equation*}

\begin{equation}\label{eq:ContFracB}
  \begin{rcases} \cfrac{1}{
            \raisebox{8.5ex}{$\displaystyle\cfrac{1}{T_{2-\delta_{j,(n-2)}}}$}
          \raisebox{8.5ex}{$\: \: -$} \begin{rcases} \cfrac{1}{\raisebox{6ex}{$\displaystyle\cfrac{1}{T_2}$}
          \raisebox{6ex}{$\: \: -$} \begin{rcases} \cfrac{1}{\raisebox{3.5ex}{$\displaystyle\cfrac{1}{T_2}$} 
          \raisebox{3.5ex}{$\: \: -$} 
          \begin{rcases} \cfrac{1}{ \hbox{\raisebox{1.5ex}{$\ddots$}} \raisebox{1.5ex}{$\: \: -$} 
          \begin{rcases} \cfrac{1}{\cfrac{1}{T_2} - T_1} \end{rcases} \text{$j=0$}
          } \end{rcases} \text{\vdots} 
          } \end{rcases} \text{$j-2$}
          } \end{rcases} \text{$j-1$}
          } \end{rcases} \text{$j$}
\end{equation}

\vspace{0.5cm} \noindent where $j = 0,1,2, ... , (n-2)$, and $\delta_{j,(n-2)}$ is the Kronecker delta function which is equal to 1 when $j=(n-2)$ and is 0 otherwise. For $j=(n-2)$, $U_{n-(j+2)}$ is equal to
\begin{equation}
U_0 = \frac{F}{k_a}.
\end{equation}

\noindent For example, for an AMM with 3 cells, we have $n=3$. This results in

\begin{align}
& & B_n = B_3 = &&  T_1 \\
j=0 \ne (n-2) & &B_{n-(j+1)} = B_2 = && \cfrac{1}{\cfrac{1}{T_2} - T_1} = \frac{T_2}{1-T_1 T_2}\\
j=1 = (n-2) & &B_{n-(j+1)} = B_1 = && \cfrac{1}{\cfrac{1}{T_1} - \cfrac{1}{\cfrac{1}{T_2} - T_1}} = \frac{T_1 (1-T_1 T_2)}{1-2 T_1 T_2}
\end{align}

\noindent which gives us the transfer functions:

\begin{equation}
\frac{U_3}{F} = \frac{1}{k_a}B_1 B_2 B_3 =  \frac{1}{k_a}\frac{T_1^2 T_2}{1-2T_1 T_2}
\end{equation}
\begin{equation}
\frac{U_2}{F} = \frac{1}{k_a}B_1 B_2 =  \frac{1}{k_a}\frac{T_1 T_2}{1-2T_1 T_2}
\end{equation}
\begin{equation}
\frac{U_1}{F} = \frac{1}{k_a}B_1  =  \frac{1}{k_a}\frac{T_1(1- T_1 T_2)}{1-2T_1 T_2}.
\end{equation}

Consequently, and as can be inferred from Figure \ref{fig:BD_R}, the transfer function relating the displacement of the $i^{th}$ cell to the force $F$ applied at the first cell can be computed using the following sequence product

\begin{equation}\label{eq:U_i/F_contfrac}
\frac{U_i}{F}= \frac{1}{k_a}\prod\limits_{j=0}^{i-1} B_{j+1}
\end{equation}

\subsection{Limiting Case: Transfer function of an infinite AMM}

\noindent An alternative form of Eq. (\ref{eq:ContFracB}), can be written as:
\begin{equation}\label{eq:B_contfrac_analytical_1}
\textit {$B_{n-(j+1)}= \frac{T_{2-\delta_{j,(n-2)}}}{1-T_{2-\delta_{j,(n-2)}} B_{n-j}}$}
\end{equation}

As the number of cells infinitely increases (i.e. $n \rightarrow \infty$), $B_{n-(j+1)}$ converges and becomes equal to its preceding value $B_{n-j}$. The converged $B_{n-(j+1)}$ and $B_{n-j}$ functions will be denoted as $\Lambda$ to distinguish them from the transfer function between two adjacent cells in a finite AMM. For an infinite chain of cells, $\delta_{j,(n-2)}$ will always be zero since all the cells can effectively be regarded as inner cells in the infinite chain. As a result, Eq. (\ref{eq:B_contfrac_analytical_1}) becomes a quadratic function which can be expressed as

\begin{equation}\label{eq:B_contfrac_analytical_2}
\textit {$T_2\Lambda^2-\Lambda+T_2 = 0$}
\end{equation}

\noindent which leads to two solutions of $\Lambda$

\begin{equation}\label{eq:B_contfrac_analytical_3}
\textit {$\Lambda_{1,2} = {\frac{1}{2 T_2}\pm \sqrt{\frac{1}{4{T_2}^2}-1}}$}
\end{equation}

\begin{figure}[h!]
\includegraphics[width=0.85\textwidth]{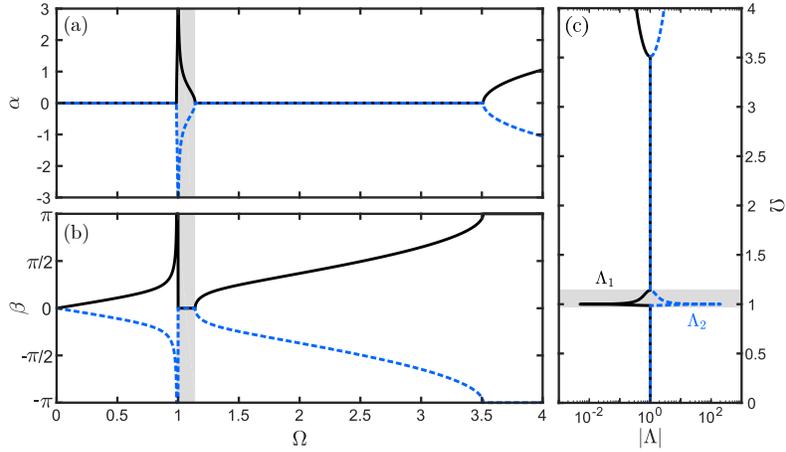}
\centering
\caption{The frequency response of (a) the attenuation constant $\alpha$ , (b) the phase difference $\beta$ and (c) the absolute value of the transfer function $|\Lambda|$ for an AMM with the same parameters used to generate Figure \ref{fig:Disp_NEM}. Any value other than unity for $|\Lambda|$ indicates a band gap (shaded areas)}
\label{fig:AbsLambda}
\end{figure}

Since $\Lambda$ represents the transfer function between two neighboring cells, it can serve as the eigenvalues of the periodic transfer matrix and hence can be written as

\begin{equation}\label{eq:prop_constant}
\textit {$\Lambda = e^{\bar{\beta}} = e^{\alpha+i\beta}$}
\end{equation}

\noindent where $\alpha$ is the attenuation constant and $\beta$ defines the phase difference between the adjacent cells \cite{AlBabaa2016a}. Consequently, the variation of $\alpha$ and $\beta$ with frequency can be determined for an infinite AMM, as shown in Figure \ref{fig:AbsLambda}. These variations perfectly coincide with the dispersion characteristics given earlier by the band structure in Figure \ref{fig:Disp_NEM} and calculated from Eqs. (\ref{eq:B_contfrac_analytical_3}) and (\ref{eq:prop_constant}), thereby confirming the limiting case of the continuous fraction transfer function model. 

\subsection{Frequency Response Functions}

The different frequency response functions (FRFs) of the AMM can be obtained by substituting $s = j\omega$, where $\omega$ is the driving frequency, and writing $T_1$ and $T_2$ in the frequency domain. Dividing throughout by $\omega_b^4$ and with a few manipulations, we obtain

\begin{equation}
T_1(\Omega) = \frac{\Gamma(1-\Omega^2)}{\Omega^4-(\Gamma+m_r+1) \Omega^2 + \Gamma}
\label{eq:FRFT1}
\end{equation}

\noindent and

\begin{equation}\label{eq:FRFT2}
\textit {$T_2(\Omega) = \frac{\Gamma(1-\Omega^2)}{\Omega^4-(2\Gamma+m_r+1) \Omega^2 + 2\Gamma}$}
\end{equation}

The frequency response function $U_i(\Omega)/F(\Omega)$ can be found from Eq. (\ref{eq:U_i/F_contfrac}) using $T_1(\Omega)$ and $T_2(\Omega)$, and thus can be expressed as

\begin{equation}
    \frac{U_i}{F} = \frac{Z_i(\Omega)}{P(\Omega)} = \frac{1}{k_a} \sum_{q=1}^{2n} \frac{a_q}{\Omega^2-\Omega_q^2}
    \label{eq:U_i/F}
\end{equation}

\noindent where $a_q$ is the residue of the $q^{th}$ pole $\Omega_q$ while $Z_i(\Omega)$ and $P(\Omega)$ are the zeros and the poles polynomials, respectively. The values of the poles $\Omega_q$ are found using Eq. (\ref{eq:analytical_poles}). The zeros polynomial $Z_i(\Omega)$ can be found from the determinant of a submatrix of Eq. (\ref{EOM_free_norm_reduced}) that result from deleting the row and the column corresponding to the excitation and sensing locations, respectively \cite{miu1993mechatronics}. This determinant represents the poles of two substructures of the system that are before and after the actuation and sensing locations, respectively, when the displacement of both locations is constrained. In our case, the forcing input is always at the outer mass of the first cell and the number of zeros sandwiched between poles will be $2(n-i)$, in addition to $i$ repeated zeros due to the presence of the local resonators. Following the scheme presented in \cite{miu1993mechatronics}, it can be shown that

\begin{equation}
    Z_i(\Omega) = \bigg |\underset{i \times i}{\Gamma \mathbf{\Omega-I}}\bigg | \bigg|\underset{n-i \times n-i}{\mathbf{\Omega^2}-\big(\Gamma \mathbf{\bar{\Psi}} + (m_r+1) \mathbf{I}\big) \mathbf{\Omega} + \Gamma \mathbf{\bar{\Psi}}}\bigg|
    \label{eq:Z_i}
\end{equation}

\noindent where $\mathbf{\bar{\Psi}} = \mathbf{\Psi}(i+1:n,i+1:n)$ is a submatrix of $\mathbf{\Psi}$. Using the same methodology described in Eqs. (\ref{EOM_free_normalized}) through (\ref{eq:eigen_vec_v}) to analytically obtain the natural frequencies, $|\mathbf{\Omega^2}-(\Gamma \mathbf{\bar{\Psi}} + (m_r+1) \mathbf{I}) \mathbf{\Omega} + \Gamma \mathbf{\bar{\Psi}}|$ can be found. As a result, $Z_i(\Omega)$ can be obtained as
\begin{equation}
    Z_i(\Omega) = \Gamma^i (\Omega^2-1)^i \prod_{k=1}^{n-i} \Omega^4_k - \big(1+m_r+2\Gamma (1-\cos \theta_k)\big)\Omega^2_k +2\Gamma (1-\cos \theta_k) 
    \label{eq:Z_i}
\end{equation}

\noindent where $\theta_k$ in this case is equal to $\frac{2k-1}{2(n-i)+1}\pi$ \cite{yueh2005eigenvalues}. Multiplying both sides of Eq. (\ref{eq:U_i/F}) by $\Omega^2-\Omega^2_q$ and evaluating the whole equation at $\Omega = \Omega_q$, the partial fraction coefficients are found to be

\begin{equation}
    a_q= \frac{Z_i(\Omega_q)}{k_a\prod_{\substack{p=1\\p\ne q}}^{2n} \Omega_q^2-\Omega^2_p}
\end{equation}

For the end-to-end transfer function $U_n/F$, the sensing location $i$ is equal to $n$ and $Z_n(\Omega)$ is simply calculated via the determinant of $[\Gamma \mathbf{\Omega-I}]_{n \times n}$, hence

\begin{equation}
    \frac{U_n}{F} = \frac{Z_n(\Omega)}{P(\Omega)} = \frac{\Gamma^n (\Omega^2-1)^n}{k_a\prod_{q=1}^{2n} \Omega^2-\Omega^2_q}
    \label{eq:U_n/F}
\end{equation}

\noindent and the partial fraction coefficients are then given by

\begin{equation}
    a_q= \frac{\Gamma^n (\Omega^2_q-1)^n}{k_a \prod_{\substack{p=1\\p\ne q}}^{2n} \Omega_q^2-\Omega^2_p}
    \label{eq:residues_n}
\end{equation}

Comparing Eqs. (\ref{eq:U_n/F}) and (\ref{eq:U_i/F}), a couple of observations can be made: 1) the term $(\Omega^2-1)$ in both numerators yields the repeated zeros at $\Omega=1$. Since $i<n$, the number of these repeated zeros for the $U_i/F$ transfer function are less than its $U_n/F$ counterpart. 2) The repeated zeros at $\Omega=1$ are the only numerator roots for $U_n/F$, while a number of anti-resonances are sandwiched in between the resonant frequencies as evident by the last term of $Z_i(\Omega)$ in Eq. (\ref{eq:Z_i}). These differences exemplify the effect of changing the sensor location on the system's transfer function, i.e., the location of the cell of which the displacement is measured along the length of the AMM. Both observations will be depicted clearly in the numerical example presented next.

\begin{figure}[h]
\centering
\includegraphics[width=\textwidth]{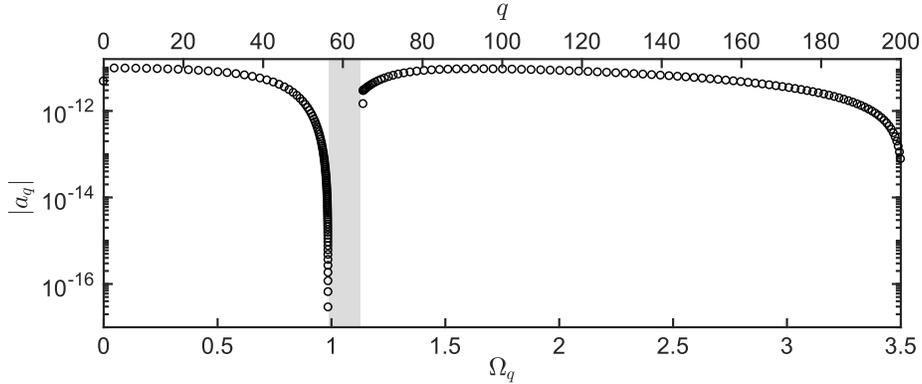}
\caption{Residues of the end-to-end transfer function (Eq. (\ref{eq:residues_n})) for an AMM of $n=100$}
\label{fig:residue_ak_n}
\end{figure}

Finally, Figure \ref{fig:residue_ak_n} shows the residues of the end-to-end transfer function from Eq. (\ref{eq:residues_n}) of an AMM of 100 unit cells. The values of the residues provide the contribution of each natural frequency to the AMM's response if the system is excited with an impulse force. It is clear that the values of the residues decrease significantly for the poles in the vicinity of the band gap (shaded region), indicating that these poles have the least contribution.

\section{\label{Numerical} Numerical Validation}
\subsection{Increasing the number of cells}

\begin{figure}[h]
\includegraphics[width=\textwidth]{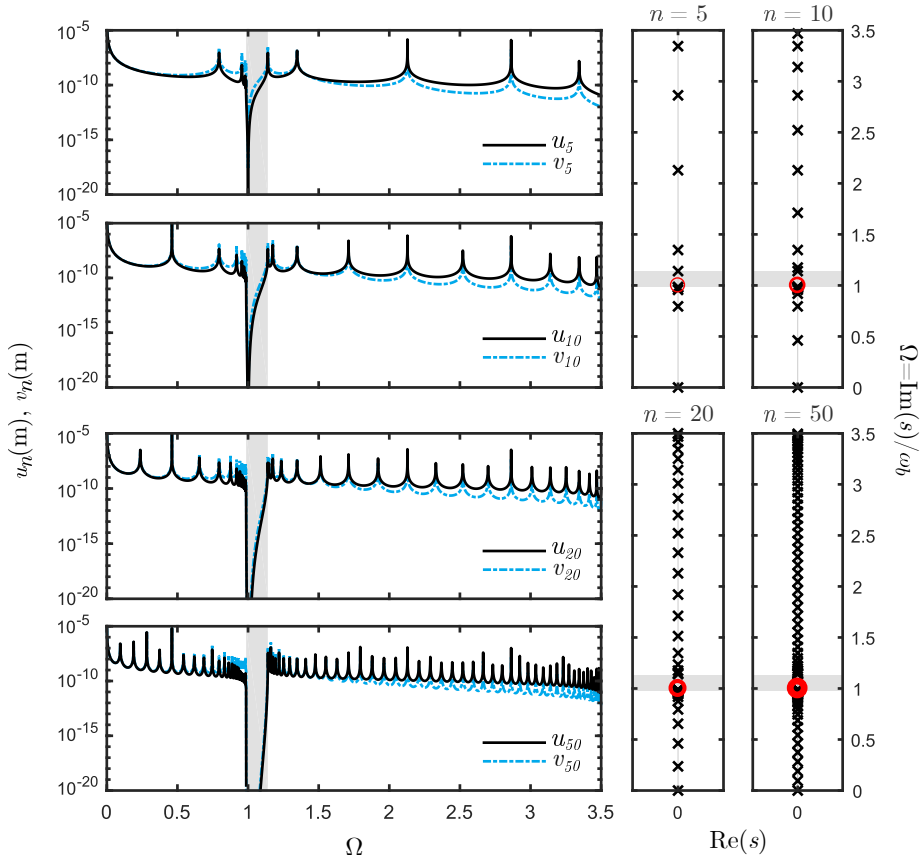}
\centering
\caption{Frequency response of the outer and inner masses of the last cell, $u_n$ and $v_n$, to an external force applied to the first cell for an AMM with $n=5$, 10, 20, and 50 cells (left column) and the corresponding pole-zero maps on the right columns. Poles and zeros are marked with crosses and circles, respectively. A thicker zero marker indicates larger algebraic multiplicity at $\Omega = 1$}
\label{fig:FRFs_diff_n}
\end{figure}

We start by considering the AMM shown in Figure \ref{fig:acoustic_mm_finite}(a) with four different number of cells: $n=5$, 10, 20 and 50. The value of the four cell parameters $m_a$, $m_b$, $k_a$, and $k_b$ are kept the same as those used earlier. Throughout this section, the focus is on the behavior of the last cell in the AMM chain in response to a force applied at the very first cell. For a sinusoidal force with a 1 N amplitude sweeping the frequency range $0<\Omega<3.5$, the frequency response of the displacements of the inner and outer masses of the last cell along with the corresponding pole-zero maps are displayed in Figure \ref{fig:FRFs_diff_n}. 

The FRFs can be directly obtained from the harmonic analysis described in Eq. (\ref{EOM_global}) or, equivalently, from the transfer function derived in  Eq. (\ref{eq:U_n/F}). At a first glance, the plots show that as the number of cells of the AMM increases, the width of the zeroed out (below $10^{-20}$) part of the response gradually increases and eventually spans the theoretically predicted band gap range of an infinite AMM, represented by the shaded region in Figure \ref{fig:FRFs_diff_n}. In other words, the AMM departs from a structure that absorbs an incident excitation at a single tuned frequency (as shown earlier in Figure \ref{fig:2cell_bode_pz}) to one that almost perfectly attenuates incident waves over a continuous frequency range, a behavior which best describes the commonly known metamaterial band gap.

The mechanism by which the emergence of a band gap behavior happens in a finite locally resonant structure with increasing the number of cells can be understood in light of the transfer functions derived in Section \ref{TFs}. The formation of the band gap is the result of three distinct phenomena that take place simultaneously in the dynamics of the AMM system. We will go over each of them separately in the next section.

\subsection{The formation mechanism of the band gap}
\subsubsection{The multiplicity effect of the locally resonant zero} 

\begin{figure}[h]
\includegraphics[width=0.4\textwidth]{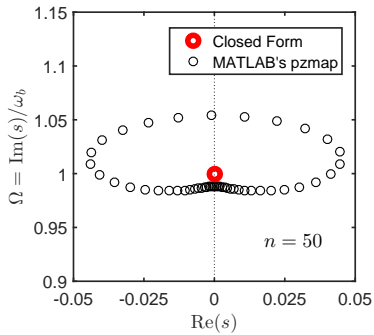}
\centering
\caption{Zeros of the transfer function $U_n/F$ of an AMM with $n=50$ as calculated using the closed-form expression in Eq. (\ref{eq:U_n/F}) and numerically using MATLAB's \textit{pzmap} operator}
\label{fig:matlab}
\end{figure}

As predicted by the derived transfer function in Eq. (\ref{eq:U_n/F}), increasing the number of cells of the AMM does not add zeros at new locations but rather increases the number of the repeated zeros at $\Omega=1$. While obvious from the numerator of the derived function, the fact that a locally resonant AMM with any number of cells $n$ has only one distinct zero location can be lost if the analysis is solely based on a numerical simulation. Figure \ref{fig:matlab} shows a common discrepancy obtained when using MATLAB's \textit{pzmap} operator to obtain the distribution of zeros for this same example. 

The figure, incorrectly, shows several zeros both above and below $\Omega=1$ with non-zero real components. Since the AMM under consideration is non-dissipative and lacks any damping elements, the roots of both the numerator and the denominator of the closed-form transfer function have to be purely imaginary (i.e. lie on the imaginary axis of the $s$-plane). The shown behavior is a result of numerical inaccuracies in the algorithm used by MATLAB to compute zeros in a system with high zero-multiplicity, which results in a cluster of approximate zeros distributed on a circle with a radius that is proportional to the machine precision \cite{matlab}. This further signifies the importance of independently deriving the closed-form expression for the AMM's transfer function using the presented block diagram approach to eventually draw accurate conclusions. 

Since the AMM's transfer function has a distinct zero location at $\Omega=1$, the response ought to be absolutely equal to zero at $\omega_b$ only. Increasing the number of repeated zeros or numerator roots (commonly known as the algebraic multiplicity) only tends to flatten out the branches of the FRF around the zero location. The degree by which the branches bend outwards increases with the increase of the multiplicity or, in this case, the number of cells. As the number of cells approaches infinity, the numerator $Z_n(\Omega)$ in Eq. (\ref{eq:U_n/F}) approaches zero for any frequency $\Omega$ as long as the value of $ \Gamma|(\Omega^2-1)|$ remains less than 1. Hence, this multiplicity effect is theoretically bounded by the range

\begin{equation}\label{eq:multiplicity_limit}
\textit {$\sqrt[]{1-\frac{1}{\Gamma}}<\Omega<\sqrt[]{1+\frac{1}{\Gamma}}$}
\end{equation}

\begin{figure}[h]
\includegraphics[width=0.85\textwidth]{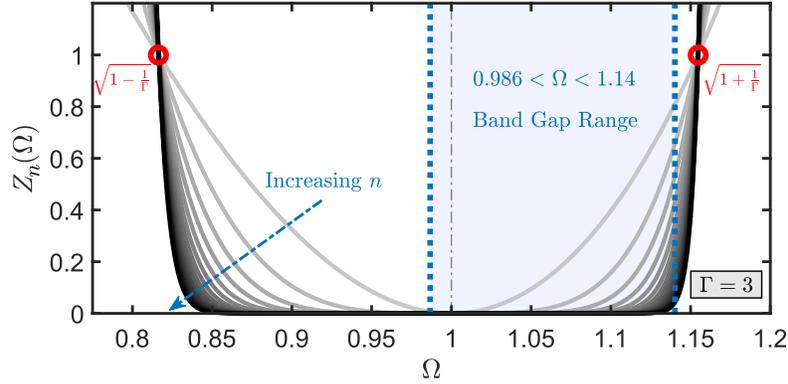}
\centering
\caption{The effect of increasing the number of repeated roots (multiplicity) of the numerator polynomial $Z_n(\Omega)$ with the increase in the number of cells $n$ in an AMM with $\Gamma=3$}
\label{fig:multiplicity}
\end{figure}

\begin{figure}[h]
\includegraphics[width=0.8\textwidth]{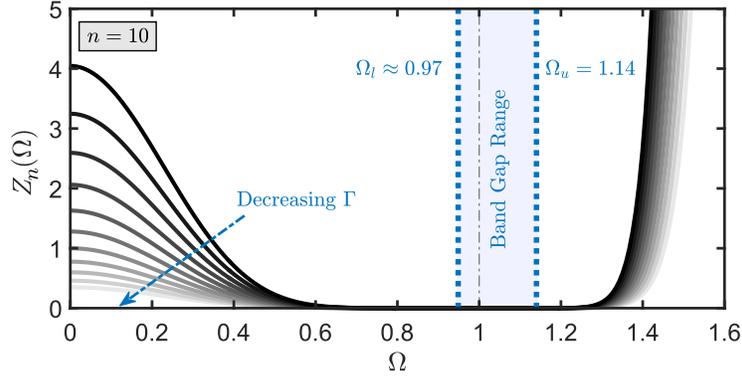}
\centering
\caption{The effect of varying $\Gamma$ on the numerator polynomial $Z_n(\Omega)$ for an AMM with a fixed number of cells ($n=10$). The mass ratio in this simulation is kept constant and equal to 0.3.  Different values of $\Gamma$ result in changing the lower band gap, which, in this case, slightly varies around $\Omega_l = 0.97$}
\label{fig:multiplicity2}
\end{figure}

Figure \ref{fig:multiplicity} shows the variation of $Z_n(\Omega)$ with the number of cells $n$ for $\Gamma = 3$. It can be seen that even though the AMM has a single zero location at $\Omega=1$, increasing $n$ gradually increases the frequency range over which $Z_n$ approaches zero via the multiplicity effect, thus widening the potential region over which the vibrations can be largely attenuated. This concept of placing multiple zeros has been exploited for robust time-delay filter design for shaping reference inputs to minimize residual vibrations of lightly damped structures~\cite{Singh93_ASME2, singh2009optimal}. This effect however, as predicted, is limited to the region bounded by Eq. (\ref{eq:multiplicity_limit}) which, in this case, corresponds to $0.816<\Omega<1.154$. Also worth noting, is the fact that all curves in Figure \ref{fig:multiplicity} with different multiplicity orders intersect at $\Omega = \sqrt[]{1-\frac{1}{\Gamma}}$ and $\Omega = \sqrt[]{1+\frac{1}{\Gamma}}$, for values of $\Gamma$ greater than 1. However, the lower bound on the multiplicity effect ceases to exist for $\Gamma<1$, as shown in Figure \ref{fig:multiplicity2}.

\subsubsection{The enforced poles at the band gap bounds}

\begin{figure}[h]
\centering
\includegraphics[width=0.6\textwidth]{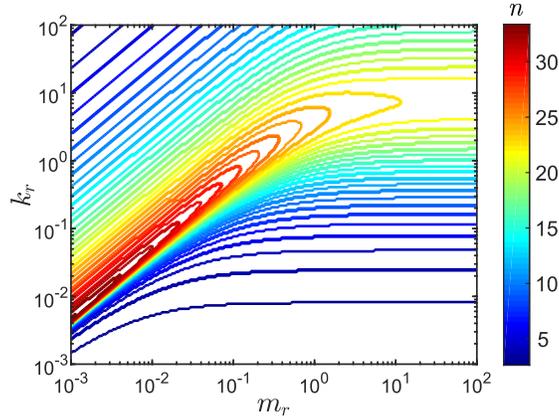}
\caption{Convergence of $\Omega_l$ to its theoretical value as a function of both $k_r$ and $m_r$}
\label{fig:omegaconv}
\end{figure}

The multiplicity range derived in Eq. (\ref{eq:multiplicity_limit}) does not match the theoretical band gap bounds derived in Eqs. (\ref{eq:Omegal}) and (\ref{eq:Omegau}), and later observed in the FRFs of Figure \ref{fig:FRFs_diff_n}(d) for a sufficiently large number of cells. In fact, Figures \ref{fig:multiplicity} and \ref{fig:multiplicity2} suggest that the band gap can extend well beyond the theoretical limits if the analysis is solely based on the transfer function zeros. The second factor contributing to the formation of the band gap, as predicted theoretically, is the location of the transfer function poles in the vicinity of $\Omega=1$. The roots of the characteristic polynomial $P(\Omega)$ of the transfer function $U_n/F$ contains a set of repeated poles at $\Omega_u=\sqrt[]{1+m_r}$, which is found by substituting $k=1$ in Eq. (\ref{eq:analytical_poles}). By knowing the value of the AMM masses, and hence $m_r$, the location of these poles can be determined and will be fixed irrespective of the number of cells $n$. The resonance peak at $\Omega_u=\sqrt[]{1+m_r}$ can be, therefore, clearly seen in all four FRF plots of Figure \ref{fig:FRFs_diff_n}. The location of these poles precisely coincides with $\Omega_u$, the upper band gap bound of the infinite AMM. This means that as the multiplicity of zeros attempts to flatten out the frequency response curve around $\Omega=1$, this effect is abruptly terminated at $\Omega_u=\sqrt[]{1+m_r}$ due to the presence of a resonant frequency of the bulk structure. The FRF responds by moving quickly from a near-zero oscillation to a state of resonance and an extremely large amplitude. The same behavior takes place at the other end of the band gap thus constituting the formation of what appears to be a zeroed response over a continuous frequency range, or the band gap. The lower bound of the band gap, $\Omega_l$, is also a pole that can be obtained from Eq. (\ref{eq:analytical_poles}) by substituting $k=n$. Unlike $\Omega_u$, this pole is not fixed and moves slowly with changing $n$. Although local resonance band gaps are commonly understood to start at $\Omega=1$, $\Omega_l$ is actually always going to be smaller than 1. In this example, for $n=50$, $\Omega_l$ is calculated to be 0.9866 to the nearest $4^{th}$ decimal place. Figure \ref{fig:omegaconv} shows the number of cells $n$ needed for $\Omega_l$ to converge as a function of both mass and stiffness ratios, $m_r$ and $k_r$, respectively. The convergence criteria is chosen such that the absolute difference between the value of the last pole in the acoustic mode (obtained from Eq. (\ref{eq:analytical_poles})) and $\Omega_l$ (obtained from Eq. (\ref{eq:Omegal})) is less than 1\%. It is seen that with $m_r>1$, the convergence tends to saturate at a specific value of $k_r$, but a larger number of cells is needed for a larger $k_r$ at a given $m_r$ value. For $m_r<1$ and $0.01<k_r<100$, both $k_r$ and $m_r$  contribute equally to $\Omega_l$ convergence. The value of $\Omega_u$ remains invariant for all values of $k_r$ and $n$ as it is only a function of $m_r$. In other words, setting $k=1$ always results in $\cos \theta_k = 1$ in Eq. (\ref{eq:analytical_poles}) which yields the solution $\Omega_{1+n}=\Omega_u$.

\vspace{-0.4cm}

\subsubsection{The absence of poles in the band gap range}

The last phenomenon that factors into the formation of the band gap in a finite AMM is the absence of resonances in the band gap frequency range. From Eq. (\ref{eq:analytical_poles}), substituting the extreme values of $\cos \theta_k$ (i.e. 1 and $-$1) provides the largest/lowest values of natural frequencies that can be obtained. The two positive roots of the analytical formula for the natural frequencies are:

\begin{equation}
    \Omega_{k_{\pm}} = \sqrt{\frac{1+m_r+2\Gamma (1-\cos \theta_k)}{2} \pm \frac{\sqrt{\big(1+m_r+2\Gamma (1-\cos \theta_k)\big)^2-8 \Gamma (1-\cos \theta_k)}}{2}}
    \label{eq:analytical_poles_roots}
\end{equation}

\noindent where $\Omega_{k_{-}}$ lies in the acoustic mode while $\Omega_{k_{+}}$ is in the optic mode. In case of $\cos \theta_k = 1$, $\Omega_{k_{\pm}} = 0, \sqrt{1+m_r}$, which are the rigid body mode and the upper bound of the band gap, as indicated earlier. If $\cos \theta_k = -1$, the two solutions are given by:

\begin{equation}
    \Omega_{k_{\pm}}= \sqrt{\frac{1+m_r+4\Gamma }{2} \pm \frac{\sqrt{(1+m_r+4\Gamma)^2-16 \Gamma}}{2}}
    \label{eq:analytical_poles_case2}
\end{equation}

\begin{figure}[h]
\includegraphics[width=\textwidth]{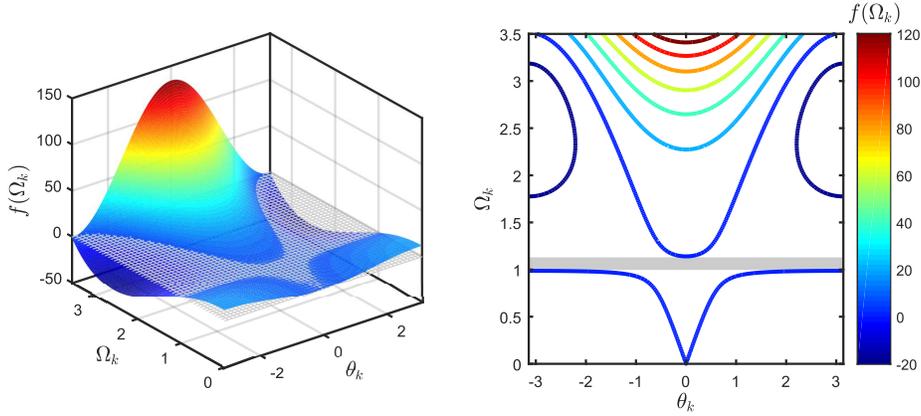}
\centering
\caption{Evaluation of the system natural frequencies $f(\Omega_k)$ from Eq. (\ref{eq:analytical_poles}) with variations of $\theta_k$ and $\Omega_k$. The meshed plane in the left column shows intersection with the zero-plane at the roots of the equation}
\label{fig:natural_frequencies}
\end{figure}

\noindent These two roots are the lower band gap limit $\Omega_{k_{-}}=\Omega_l$ and the limit at which the unbounded \textit{stop band} starts \cite{nouh2017spatial}. This condition of $\cos \theta_k =-1$ is, however, not possible since it requires $k$ (which has a maximum possible value of $n$) to be equal to $n+1$. Hence, a finite AMM will have natural frequencies in the proximity of these two roots as the number of cells increases. Combining all cases shows that the possible natural frequencies are within two ranges:

\begin{equation}
    0\leq\Omega_{k_{-}}<\Omega_l
    \label{eq:analytical_poles_limits1}
\end{equation}

\begin{equation}
\Omega_u\leq\Omega_{k_{+}}<\sqrt{\frac{1+m_r+4\Gamma }{2} + \frac{\sqrt{(1+m_r+4\Gamma)^2-16 \Gamma}}{2} } \label{eq:analytical_poles_limits2}
\end{equation}

\noindent which do not lie within the band gap region. Figure \ref{fig:natural_frequencies} graphically illustrates the regions where Eq. (\ref{eq:analytical_poles}) crosses the zero-plane. As predicted, there are no solutions within the band gap region.

\begin{figure}[h]
\includegraphics[width=\textwidth]{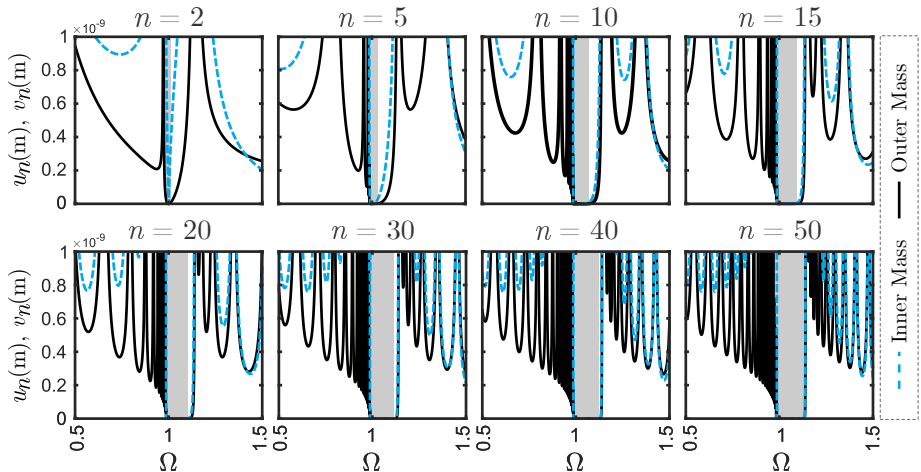}
\centering
\caption{Evolution of the local resonance band gap in finite metamaterials from a 2-cell AMM upto a 50-cell AMM shown from left top corner to the right bottom corner. The shaded area shows the portion of FRF below $10^{-20}$.}
\label{fig:evolution}
\end{figure}

The pole-zero maps in Figure \ref{fig:FRFs_diff_n} interestingly show that, irrespective of the number of cells $n$ that constitute the AMM, the overall system poles are equally split around the  band gap range with no poles lying in the shaded region. In the previous section, it was explained how the presence of two resonance peaks flanking the repeated zeros at $\Omega=1$, define the bandwidth of the band gap. This last phenomenon is, therefore, essential to ensure that the multiplicity effect which contributes to extending the effect of the single zero location over the entire band gap range is not interrupted within the band gap. Finally, by adopting a linear scale and a close-up of the FRFs, Figure \ref{fig:evolution} nicely illustrates the evolution of the band gap from a single anti-resonance (for an AMM with $n=2$) to a continuous attenuation over an extended range ($n=50$).

\vspace{-0.4cm}

\subsection{\label{sensor}Effect of sensor location}

\begin{figure}[h!]
\includegraphics[width=\textwidth]{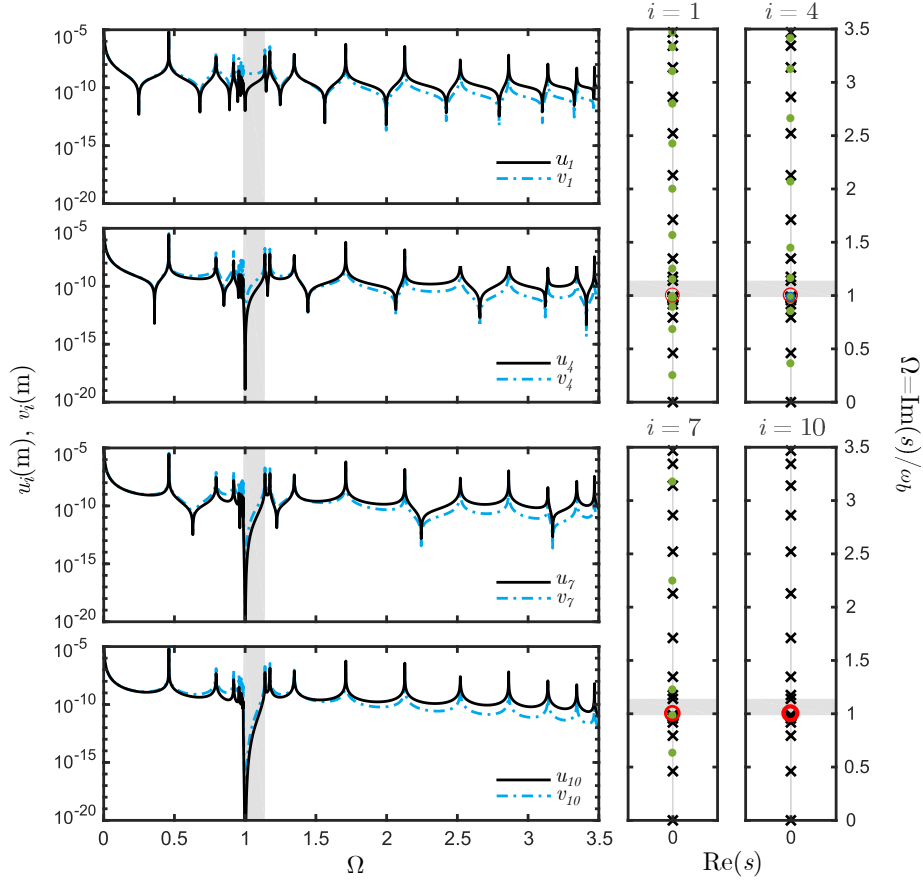}
\centering
\caption{Frequency response of the outer and inner masses, $u_i$ and $v_i$, of the 1st, 4th, 7th, and 10th (last) cell in an AMM with $n=10$ cells (left column) along with their corresponding pole-zero maps (right columns). The red circles represents the zeros located at $\Omega=1$, while the green dots are for the zeros caused from moving the measurement location. The poles are represented as black crosses.}
\label{fig:FRFs_diff_loc}
\end{figure}

In this section, we examine the spatial attenuation of the propagating wave from the excitation location at one end of the AMM to different locations along the length of the AMM. Hence, we shift our focus to the transfer function $U_i/F$ derived in Eq. (\ref{eq:U_i/F}). For an AMM with $n=10$ cells and the same values for the cell parameters $m_a$, $m_b$, $k_a$, and $k_b$, Figure \ref{fig:FRFs_diff_loc} shows the frequency responses of the \(1^{st}\), \(4^{th}\), \(7^{th}\) and \(10^{th}\) (and last) cells, respectively, to an external force acting on the outer mass of the \(1^{st}\) cell and the corresponding pole-zero maps. 

The FRFs shown in Figure \ref{fig:FRFs_diff_loc} show that the degree of attenuation in the AMM increases as we move further away from the source, i.e., as $i$ approaches $n$. From a transfer function perspective, the order of zero multiplicity (number of repeated zeros at $\Omega=1$) is the same order as $i$, as captured by the $\Gamma^i(\Omega^2-1)^i$ term in the numerator $Z_i(\Omega)$ of Eq. (\ref{eq:Z_i}). As the sensor is moved closer to the exciting force (i.e. as $i$ approaches 1), some of these repeated zeros migrate to locations sandwiched between the system's poles. For a perfectly collocated sensor an actuator, each set of two neighboring poles will be separated by at least one zero. This zero migration behavior has been highlighted in literature in the context of controlled flexible structures \cite{Martin78, Wie1981}.

\section{Conclusions}

This paper has presented an in-depth analysis of the formation of local resonance band gaps in locally resonant acoustic metamaterials of a finite length. A block diagram reduction approach was used to derive closed-form expressions for a lumped AMM with any number of cells $n$ as well as at any cell location $i$. By deriving closed-form expressions for the force to end-displacement transfer functions, the frequency response of an AMM with a given set of parameters was obtained. The presented framework explained how the evolution of the band gap from a single anti-resonance to an attenuation of incident excitations over a continuous frequency range. The results are matched with the theoretical bounds of the band gap as derived by the commonly adopted Bloch-wave model and the dispersion relations for a traveling wave in an infinitely-long metamaterial. The emergence of the local resonance band gap was explained using the derived system dynamics in light of three separate, yet simultaneous, phenomenon:

\begin{enumerate}
\item The existence of an $n^{th}$ order multiplicity of repeated zeros at the distinct location of the local resonator $\omega_b$ which flattens out the FRF curves of the AMM's displacement, thus zeroing out the response around $\Omega=1$.
\item The existence of system poles at both ends of the band gap that breaks the multiplicity effect. The pole north of the band gap is a fixed resonance at $\Omega_u$ that is independent of the number of cells and is solely a function of the mass ratio, while that south of the band gap varies with the AMM size and approaches its theoretical counterpart at sufficiently large values of $n$.
\item The distribution of poles on both sides of the band gap giving rise to different acoustic and optic resonant modes with the absence of any other poles in the band gap region.
\end{enumerate}

The presented results facilitate the understanding of this type of local resonance band gaps in finite structures from a vibrations, rather than a wave-propagation standpoint. The study bridges the gap between the dispersion of waves in theoretical infinite structures and the structural dynamics of a finite, and thus physically realizable, locally resonant metamaterial. Finally, investigating AMMs in the context of frequency domain tools and PZ-maps sets a future framework for implementing robust control techniques.


\bibliography{mybibfile}

\begin{thebibliography}{48}
\providecommand{\natexlab}[1]{#1}
\providecommand{\url}[1]{\texttt{#1}}
\providecommand{\href}[2]{#2}
\providecommand{\path}[1]{#1}
\providecommand{\eprint}[1]{\href{http://arxiv.org/abs/#1}{\path{#1}}}
\providecommand{\DOIprefix}{doi:}
\providecommand{\ArXivprefix}{arXiv:}
\providecommand{\URLprefix}{URL: }
\providecommand{\Pubmedprefix}{pmid:}
\providecommand{\doi}[1]{\href{http://dx.doi.org/#1}{\path{#1}}}
\providecommand{\Pubmed}[1]{\href{pmid:#1}{\path{#1}}}
\providecommand{\BIBand}{and}
\providecommand{\bibinfo}[2]{#2}
\ifx\xfnm\undefined \def\xfnm[#1]{\unskip,\space#1}\fi
\bibitem[{Liu et~al.(2000)Liu, Zhang, Mao, Zhu, Yang, Chan et~al.}]{liu_sonic}
\bibinfo{author}{Liu\xfnm[ Z.]}, \bibinfo{author}{Zhang\xfnm[ X.]},
  \bibinfo{author}{Mao\xfnm[ Y.]}, \bibinfo{author}{Zhu\xfnm[ Y.]},
  \bibinfo{author}{Yang\xfnm[ Z.]}, \bibinfo{author}{Chan\xfnm[ C.T.]}, et~al.
\newblock \bibinfo{title}{Locally resonant sonic materials}.
\newblock \bibinfo{journal}{Science}
  \bibinfo{year}{2000};\bibinfo{volume}{289}(\bibinfo{number}{5485}):\bibinfo{pages}{1734--1736}.
\bibitem[{Huang and Sun(2011)}]{huang2011study}
\bibinfo{author}{Huang\xfnm[ H.H.]}, \bibinfo{author}{Sun\xfnm[ C.T.]}.
\newblock \bibinfo{title}{{A study of band-gap phenomena of two locally
  resonant acoustic metamaterials}}.
\newblock \bibinfo{journal}{Proceedings of the Institution of Mechanical
  Engineers, Part N: Journal of Nanoengineering and Nanosystems}
  \bibinfo{year}{2011};:\bibinfo{pages}{1740349911409981}.
\bibitem[{Huang and Sun(2010)}]{Huang2010}
\bibinfo{author}{Huang\xfnm[ G.L.]}, \bibinfo{author}{Sun\xfnm[ C.T.]}.
\newblock \bibinfo{title}{{Band Gaps in a Multiresonator Acoustic
  Metamaterial}}.
\newblock \bibinfo{journal}{Journal of Vibration and Acoustics}
  \bibinfo{year}{2010};\bibinfo{volume}{132}(\bibinfo{number}{3}):\bibinfo{pages}{031003}.
\newblock \DOIprefix\doi{10.1115/1.4000784}.
\bibitem[{Wang et~al.(2004)Wang, Yu, Wen, Liu and Wen}]{wang2004}
\bibinfo{author}{Wang\xfnm[ G.]}, \bibinfo{author}{Yu\xfnm[ D.]},
  \bibinfo{author}{Wen\xfnm[ J.]}, \bibinfo{author}{Liu\xfnm[ Y.]},
  \bibinfo{author}{Wen\xfnm[ X.]}.
\newblock \bibinfo{title}{One-dimensional phononic crystals with locally
  resonant structures}.
\newblock \bibinfo{journal}{Physics Letters A}
  \bibinfo{year}{2004};\bibinfo{volume}{327}(\bibinfo{number}{5}):\bibinfo{pages}{512--521}.
\bibitem[{Xiao et~al.(2012)Xiao, Wen and Wen}]{xiao2012longitudinal}
\bibinfo{author}{Xiao\xfnm[ Y.]}, \bibinfo{author}{Wen\xfnm[ J.]},
  \bibinfo{author}{Wen\xfnm[ X.]}.
\newblock \bibinfo{title}{Longitudinal wave band gaps in metamaterial-based
  elastic rods containing multi-degree-of-freedom resonators}.
\newblock \bibinfo{journal}{New Journal of Physics}
  \bibinfo{year}{2012};\bibinfo{volume}{14}(\bibinfo{number}{3}):\bibinfo{pages}{033042}.
\bibitem[{Yu et~al.(2006{\natexlab{a}})Yu, Liu, Zhao, Wang and Qiu}]{yu2006a}
\bibinfo{author}{Yu\xfnm[ D.]}, \bibinfo{author}{Liu\xfnm[ Y.]},
  \bibinfo{author}{Zhao\xfnm[ H.]}, \bibinfo{author}{Wang\xfnm[ G.]},
  \bibinfo{author}{Qiu\xfnm[ J.]}.
\newblock \bibinfo{title}{Flexural vibration band gaps in euler-bernoulli beams
  with locally resonant structures with two degrees of freedom}.
\newblock \bibinfo{journal}{Physical Review B}
  \bibinfo{year}{2006}{\natexlab{a}};\bibinfo{volume}{73}(\bibinfo{number}{6}):\bibinfo{pages}{064301}.
\bibitem[{Yu et~al.(2006{\natexlab{b}})Yu, Liu, Wang, Zhao and Qiu}]{yu2006b}
\bibinfo{author}{Yu\xfnm[ D.]}, \bibinfo{author}{Liu\xfnm[ Y.]},
  \bibinfo{author}{Wang\xfnm[ G.]}, \bibinfo{author}{Zhao\xfnm[ H.]},
  \bibinfo{author}{Qiu\xfnm[ J.]}.
\newblock \bibinfo{title}{Flexural vibration band gaps in timoshenko beams with
  locally resonant structures}.
\newblock \bibinfo{journal}{Journal of Applied Physics}
  \bibinfo{year}{2006}{\natexlab{b}};\bibinfo{volume}{100}(\bibinfo{number}{12}):\bibinfo{pages}{124901}.
\bibitem[{Sun et~al.(2010)Sun, Du and Pai}]{sun2010}
\bibinfo{author}{Sun\xfnm[ H.]}, \bibinfo{author}{Du\xfnm[ X.]},
  \bibinfo{author}{Pai\xfnm[ P.F.]}.
\newblock \bibinfo{title}{Theory of metamaterial beams for broadband vibration
  absorption}.
\newblock \bibinfo{journal}{Journal of Intelligent Material Systems and
  Structures} \bibinfo{year}{2010};.
\bibitem[{Nouh et~al.(2014)Nouh, Aldraihem and Baz}]{Nouh2014}
\bibinfo{author}{Nouh\xfnm[ M.]}, \bibinfo{author}{Aldraihem\xfnm[ O.]},
  \bibinfo{author}{Baz\xfnm[ A.]}.
\newblock \bibinfo{title}{{Vibration Characteristics of Metamaterial Beams With
  Periodic Local Resonances}}.
\newblock \bibinfo{journal}{Journal of Vibration and Acoustics}
  \bibinfo{year}{2014};\bibinfo{volume}{136}(\bibinfo{number}{6}):\bibinfo{pages}{61012}.
\newblock \DOIprefix\doi{10.1115/1.4028453}.
\bibitem[{Pai et~al.(2014)Pai, Peng and Jiang}]{Pai2014}
\bibinfo{author}{Pai\xfnm[ P.F.]}, \bibinfo{author}{Peng\xfnm[ H.]},
  \bibinfo{author}{Jiang\xfnm[ S.]}.
\newblock \bibinfo{title}{{Acoustic metamaterial beams based on multi-frequency
  vibration absorbers}}.
\newblock \bibinfo{journal}{International Journal of Mechanical Sciences}
  \bibinfo{year}{2014};\bibinfo{volume}{79}:\bibinfo{pages}{195--205}.
\bibitem[{Khajehtourian and Hussein(2014)}]{hussein_AIP}
\bibinfo{author}{Khajehtourian\xfnm[ R.]}, \bibinfo{author}{Hussein\xfnm[ M.]}.
\newblock \bibinfo{title}{Dispersion characteristics of a nonlinear elastic
  metamaterial}.
\newblock \bibinfo{journal}{AIP Advances}
  \bibinfo{year}{2014};\bibinfo{volume}{4}(\bibinfo{number}{12}):\bibinfo{pages}{124308}.
\bibitem[{Xiao et~al.(2013)Xiao, Wen, Yu and Wen}]{xiao2013flexural}
\bibinfo{author}{Xiao\xfnm[ Y.]}, \bibinfo{author}{Wen\xfnm[ J.]},
  \bibinfo{author}{Yu\xfnm[ D.]}, \bibinfo{author}{Wen\xfnm[ X.]}.
\newblock \bibinfo{title}{Flexural wave propagation in beams with periodically
  attached vibration absorbers: band-gap behavior and band formation
  mechanisms}.
\newblock \bibinfo{journal}{Journal of Sound and Vibration}
  \bibinfo{year}{2013};\bibinfo{volume}{332}(\bibinfo{number}{4}):\bibinfo{pages}{867--893}.
\bibitem[{Baravelli and Ruzzene(2013)}]{Baravelli2013}
\bibinfo{author}{Baravelli\xfnm[ E.]}, \bibinfo{author}{Ruzzene\xfnm[ M.]}.
\newblock \bibinfo{title}{{Internally resonating lattices for bandgap
  generation and low-frequency vibration control}}.
\newblock \bibinfo{journal}{Journal of Sound and Vibration}
  \bibinfo{year}{2013};\bibinfo{volume}{332}(\bibinfo{number}{25}):\bibinfo{pages}{6562--6579}.
\newblock \DOIprefix\doi{10.1016/j.jsv.2013.08.014}.
\bibitem[{Zhu et~al.(2014)Zhu, Liu, Hu, Sun and Huang}]{Zhu2014}
\bibinfo{author}{Zhu\xfnm[ R.]}, \bibinfo{author}{Liu\xfnm[ X.]},
  \bibinfo{author}{Hu\xfnm[ G.]}, \bibinfo{author}{Sun\xfnm[ C.]},
  \bibinfo{author}{Huang\xfnm[ G.]}.
\newblock \bibinfo{title}{{A chiral elastic metamaterial beam for broadband
  vibration suppression}}.
\newblock \bibinfo{journal}{Journal of Sound and Vibration}
  \bibinfo{year}{2014};\bibinfo{volume}{333}(\bibinfo{number}{10}):\bibinfo{pages}{2759--2773}.
\newblock \DOIprefix\doi{10.1016/j.jsv.2014.01.009}.
\bibitem[{Krushynska et~al.(2014)Krushynska, Kouznetsova and Geers}]{towards}
\bibinfo{author}{Krushynska\xfnm[ A.]}, \bibinfo{author}{Kouznetsova\xfnm[
  V.]}, \bibinfo{author}{Geers\xfnm[ M.]}.
\newblock \bibinfo{title}{Towards optimal design of locally resonant acoustic
  metamaterials}.
\newblock \bibinfo{journal}{Journal of the Mechanics and Physics of Solids}
  \bibinfo{year}{2014};\bibinfo{volume}{71}:\bibinfo{pages}{179--196}.
\bibitem[{Peng and Pai(2014)}]{pai3}
\bibinfo{author}{Peng\xfnm[ H.]}, \bibinfo{author}{Pai\xfnm[ P.F.]}.
\newblock \bibinfo{title}{Acoustic metamaterial plates for elastic wave
  absorption and structural vibration suppression}.
\newblock \bibinfo{journal}{International Journal of Mechanical Sciences}
  \bibinfo{year}{2014};\bibinfo{volume}{89}:\bibinfo{pages}{350--361}.
\bibitem[{Wang and Wang(2013)}]{wang2013}
\bibinfo{author}{Wang\xfnm[ Y.]}, \bibinfo{author}{Wang\xfnm[ Y.]}.
\newblock \bibinfo{title}{Complete bandgaps in two-dimensional phononic crystal
  slabs with resonators}.
\newblock \bibinfo{journal}{Journal of Applied Physics}
  \bibinfo{year}{2013};\bibinfo{volume}{114}(\bibinfo{number}{4}):\bibinfo{pages}{043509}.
\bibitem[{Nouh et~al.(2015)Nouh, Aldraihem and Baz}]{Nouh2015}
\bibinfo{author}{Nouh\xfnm[ M.]}, \bibinfo{author}{Aldraihem\xfnm[ O.]},
  \bibinfo{author}{Baz\xfnm[ A.]}.
\newblock \bibinfo{title}{{Wave propagation in metamaterial plates with
  periodic local resonances}}.
\newblock \bibinfo{journal}{Journal of Sound and Vibration}
  \bibinfo{year}{2015};\bibinfo{volume}{341}:\bibinfo{pages}{53--73}.
\newblock \DOIprefix\doi{10.1016/j.jsv.2014.12.030}.
\bibitem[{Gonella et~al.(2009)Gonella, To and Liu}]{Gonella2009}
\bibinfo{author}{Gonella\xfnm[ S.]}, \bibinfo{author}{To\xfnm[ A.C.]},
  \bibinfo{author}{Liu\xfnm[ W.K.]}.
\newblock \bibinfo{title}{{Interplay between phononic bandgaps and
  piezoelectric microstructures for energy harvesting}}.
\newblock \bibinfo{journal}{Journal of the Mechanics and Physics of Solids}
  \bibinfo{year}{2009};\bibinfo{volume}{57}(\bibinfo{number}{3}):\bibinfo{pages}{621--633}.
\newblock \DOIprefix\doi{10.1016/j.jmps.2008.11.002}.
\bibitem[{Celli and Gonella(2015)}]{Celli2015}
\bibinfo{author}{Celli\xfnm[ P.]}, \bibinfo{author}{Gonella\xfnm[ S.]}.
\newblock \bibinfo{title}{{Tunable directivity in metamaterials with
  reconfigurable cell symmetry}}.
\newblock \bibinfo{journal}{Applied Physics Letters}
  \bibinfo{year}{2015};\bibinfo{volume}{106}(\bibinfo{number}{9}).
\newblock \DOIprefix\doi{10.1063/1.4914011}.
\bibitem[{Chen et~al.(2016{\natexlab{a}})Chen, Hu and Huang}]{Chen2015}
\bibinfo{author}{Chen\xfnm[ Y.]}, \bibinfo{author}{Hu\xfnm[ J.]},
  \bibinfo{author}{Huang\xfnm[ G.]}.
\newblock \bibinfo{title}{{A design of active elastic metamaterials for control
  of flexural waves using the transformation method}}.
\newblock \bibinfo{journal}{Journal of Intelligent Material Systems and
  Structures}
  \bibinfo{year}{2016}{\natexlab{a}};\bibinfo{volume}{27}(\bibinfo{number}{10}):\bibinfo{pages}{1337--1347}.
\newblock \DOIprefix\doi{10.1177/1045389X15590273}.
\bibitem[{Chen et~al.(2014)Chen, Huang and Sun}]{chen2014piezo}
\bibinfo{author}{Chen\xfnm[ Y.]}, \bibinfo{author}{Huang\xfnm[ G.]},
  \bibinfo{author}{Sun\xfnm[ C.]}.
\newblock \bibinfo{title}{Band gap control in an active elastic metamaterial
  with negative capacitance piezoelectric shunting}.
\newblock \bibinfo{journal}{Journal of Vibration and Acoustics}
  \bibinfo{year}{2014};\bibinfo{volume}{136}(\bibinfo{number}{6}):\bibinfo{pages}{061008}.
\bibitem[{Nouh et~al.(2016)Nouh, Aldraihem and Baz}]{Nouh2016}
\bibinfo{author}{Nouh\xfnm[ M.]}, \bibinfo{author}{Aldraihem\xfnm[ O.]},
  \bibinfo{author}{Baz\xfnm[ A.]}.
\newblock \bibinfo{title}{Periodic metamaterial plates with smart tunable local
  resonators}.
\newblock \bibinfo{journal}{Journal of Intelligent Material Systems and
  Structures}
  \bibinfo{year}{2016};\bibinfo{volume}{27}(\bibinfo{number}{13}):\bibinfo{pages}{1829--1845}.
\bibitem[{Bloch(1929)}]{Bloch1929}
\bibinfo{author}{Bloch\xfnm[ F.]}.
\newblock \bibinfo{title}{{\"U}ber die quantenmechanik der elektronen in
  kristallgittern}.
\newblock \bibinfo{journal}{Zeitschrift f{\"u}r physik}
  \bibinfo{year}{1929};\bibinfo{volume}{52}(\bibinfo{number}{7-8}):\bibinfo{pages}{555--600}.
\bibitem[{Hussein et~al.(2014)Hussein, Leamy and Ruzzene}]{Hussein2014}
\bibinfo{author}{Hussein\xfnm[ M.I.]}, \bibinfo{author}{Leamy\xfnm[ M.J.]},
  \bibinfo{author}{Ruzzene\xfnm[ M.]}.
\newblock \bibinfo{title}{{Dynamics of Phononic Materials and Structures:
  Historical Origins, Recent Progress, and Future Outlook}}.
\newblock \bibinfo{journal}{Applied Mechanics Reviews}
  \bibinfo{year}{2014};\bibinfo{volume}{66}(\bibinfo{number}{4}):\bibinfo{pages}{040802}.
\newblock \DOIprefix\doi{10.1115/1.4026911}.
\bibitem[{Mead(1970)}]{Mead1970}
\bibinfo{author}{Mead\xfnm[ D.]}.
\newblock \bibinfo{title}{{Free wave propagation in periodically supported,
  infinite beams}}.
\newblock \bibinfo{journal}{Journal of Sound and Vibration}
  \bibinfo{year}{1970};\bibinfo{volume}{11}(\bibinfo{number}{2}):\bibinfo{pages}{181--197}.
\newblock \DOIprefix\doi{10.1016/S0022-460X(70)80062-1}.
\bibitem[{Mead(1971)}]{Mead1971}
\bibinfo{author}{Mead\xfnm[ D.J.]}.
\newblock \bibinfo{title}{{Vibration Response and Wave Propagation in Periodic
  Structures}}.
\newblock \bibinfo{journal}{Journal of Engineering for Industry}
  \bibinfo{year}{1971};\bibinfo{volume}{93}(\bibinfo{number}{3}):\bibinfo{pages}{783}.
\newblock \DOIprefix\doi{10.1115/1.3428014}.
\bibitem[{Faulkner and Hong(1985)}]{Faulkner1985}
\bibinfo{author}{Faulkner\xfnm[ M.G.]}, \bibinfo{author}{Hong\xfnm[ D.P.]}.
\newblock \bibinfo{title}{{Free vibrations of a mono-coupled periodic system}}.
\newblock \bibinfo{journal}{Journal of Sound and Vibration}
  \bibinfo{year}{1985};\bibinfo{volume}{99}(\bibinfo{number}{1}):\bibinfo{pages}{29--42}.
\newblock \DOIprefix\doi{10.1016/0022-460X(85)90443-2}.
\bibitem[{Huang et~al.(2009)Huang, Sun and Huang}]{huang2009negative}
\bibinfo{author}{Huang\xfnm[ H.H.]}, \bibinfo{author}{Sun\xfnm[ C.T.]},
  \bibinfo{author}{Huang\xfnm[ G.L.]}.
\newblock \bibinfo{title}{On the negative effective mass density in acoustic
  metamaterials}.
\newblock \bibinfo{journal}{International Journal of Engineering Science}
  \bibinfo{year}{2009};\bibinfo{volume}{47}(\bibinfo{number}{4}):\bibinfo{pages}{610--617}.
\bibitem[{Pai(2010)}]{Pai2010}
\bibinfo{author}{Pai\xfnm[ P.F.]}.
\newblock \bibinfo{title}{{Metamaterial-based Broadband Elastic Wave
  Absorber}}.
\newblock \bibinfo{journal}{Journal of Intelligent Material Systems and
  Structures}
  \bibinfo{year}{2010};\bibinfo{volume}{21}(\bibinfo{number}{5}):\bibinfo{pages}{517--528}.
\newblock \DOIprefix\doi{10.1177/1045389X09359436}.
\bibitem[{Gupta(1970)}]{gupta1970natural}
\bibinfo{author}{Gupta\xfnm[ G.S.]}.
\newblock \bibinfo{title}{Natural flexural waves and the normal modes of
  periodically-supported beams and plates}.
\newblock \bibinfo{journal}{Journal of Sound and Vibration}
  \bibinfo{year}{1970};\bibinfo{volume}{13}(\bibinfo{number}{1}):\bibinfo{pages}{89--101}.
\bibitem[{Nielsen and Sorokin(2015)}]{nielsen2015periodicity}
\bibinfo{author}{Nielsen\xfnm[ R.]}, \bibinfo{author}{Sorokin\xfnm[ S.]}.
\newblock \bibinfo{title}{Periodicity effects of axial waves in elastic
  compound rods}.
\newblock \bibinfo{journal}{Journal of Sound and Vibration}
  \bibinfo{year}{2015};\bibinfo{volume}{353}:\bibinfo{pages}{135--149}.
\bibitem[{Hvatov and Sorokin(2015)}]{hvatov2015free}
\bibinfo{author}{Hvatov\xfnm[ A.]}, \bibinfo{author}{Sorokin\xfnm[ S.]}.
\newblock \bibinfo{title}{Free vibrations of finite periodic structures in
  pass-and stop-bands of the counterpart infinite waveguides}.
\newblock \bibinfo{journal}{Journal of Sound and Vibration}
  \bibinfo{year}{2015};\bibinfo{volume}{347}:\bibinfo{pages}{200--217}.
\bibitem[{Sugino et~al.(2016)Sugino, Leadenham, Ruzzene and
  Erturk}]{sugino2016mechanism}
\bibinfo{author}{Sugino\xfnm[ C.]}, \bibinfo{author}{Leadenham\xfnm[ S.]},
  \bibinfo{author}{Ruzzene\xfnm[ M.]}, \bibinfo{author}{Erturk\xfnm[ A.]}.
\newblock \bibinfo{title}{On the mechanism of bandgap formation in locally
  resonant finite elastic metamaterials}.
\newblock \bibinfo{journal}{Journal of Applied Physics}
  \bibinfo{year}{2016};\bibinfo{volume}{120}(\bibinfo{number}{13}):\bibinfo{pages}{134501}.
\bibitem[{Hussein and Frazier(2013)}]{hussein_damp1}
\bibinfo{author}{Hussein\xfnm[ M.I.]}, \bibinfo{author}{Frazier\xfnm[ M.J.]}.
\newblock \bibinfo{title}{Metadamping: An emergent phenomenon in dissipative
  metamaterials}.
\newblock \bibinfo{journal}{Journal of Sound and Vibration}
  \bibinfo{year}{2013};\bibinfo{volume}{332}(\bibinfo{number}{20}):\bibinfo{pages}{4767--4774}.
\bibitem[{Frazier and Hussein(2016)}]{hussein_damp2}
\bibinfo{author}{Frazier\xfnm[ M.J.]}, \bibinfo{author}{Hussein\xfnm[ M.I.]}.
\newblock \bibinfo{title}{Generalized bloch's theorem for viscous
  metamaterials: Dispersion and effective properties based on frequencies and
  wavenumbers that are simultaneously complex}.
\newblock \bibinfo{journal}{Comptes Rendus Physique}
  \bibinfo{year}{2016};\bibinfo{volume}{17}(\bibinfo{number}{5}):\bibinfo{pages}{565--577}.
\bibitem[{Chen et~al.(2016{\natexlab{b}})Chen, Barnhart, Chen, Hu, Sun and
  Huang}]{huang_damp3}
\bibinfo{author}{Chen\xfnm[ Y.]}, \bibinfo{author}{Barnhart\xfnm[ M.]},
  \bibinfo{author}{Chen\xfnm[ J.]}, \bibinfo{author}{Hu\xfnm[ G.]},
  \bibinfo{author}{Sun\xfnm[ C.]}, \bibinfo{author}{Huang\xfnm[ G.]}.
\newblock \bibinfo{title}{Dissipative elastic metamaterials for broadband wave
  mitigation at subwavelength scale}.
\newblock \bibinfo{journal}{Composite Structures}
  \bibinfo{year}{2016}{\natexlab{b}};\bibinfo{volume}{136}:\bibinfo{pages}{358--371}.
\bibitem[{Andreassen and Jensen(2013)}]{andreassen_damp4}
\bibinfo{author}{Andreassen\xfnm[ E.]}, \bibinfo{author}{Jensen\xfnm[ J.S.]}.
\newblock \bibinfo{title}{Analysis of phononic bandgap structures with
  dissipation}.
\newblock \bibinfo{journal}{Journal of Vibration and Acoustics}
  \bibinfo{year}{2013};\bibinfo{volume}{135}(\bibinfo{number}{4}):\bibinfo{pages}{041015}.
\bibitem[{Yueh(2005)}]{yueh2005eigenvalues}
\bibinfo{author}{Yueh\xfnm[ W.C.]}.
\newblock \bibinfo{title}{Eigenvalues of several tridiagonal matrices}.
\newblock \bibinfo{journal}{Applied mathematics e-notes}
  \bibinfo{year}{2005};\bibinfo{volume}{5}(\bibinfo{number}{66-74}):\bibinfo{pages}{210--230}.
\bibitem[{Hensley(2006)}]{hensley_contfrac}
\bibinfo{author}{Hensley\xfnm[ D.]}.
\newblock \bibinfo{title}{Continued fractions}; vol.~\bibinfo{volume}{20}.
\newblock \bibinfo{publisher}{World Scientific}; \bibinfo{year}{2006}.
\bibitem[{Al-Ba'ba'a and Nouh(2017)}]{AlBabaa2016a}
\bibinfo{author}{Al-Ba'ba'a\xfnm[ H.]}, \bibinfo{author}{Nouh\xfnm[ M.]}.
\newblock \bibinfo{title}{An investigation of vibrational power flow in
  one-dimensional dissipative phononic structures}.
\newblock \bibinfo{journal}{Journal of Vibration and Acoustics}
  \bibinfo{year}{2017};\bibinfo{volume}{139}(\bibinfo{number}{2}):\bibinfo{pages}{021003}.
\bibitem[{Miu(1993)}]{miu1993mechatronics}
\bibinfo{author}{Miu\xfnm[ D.]}.
\newblock \bibinfo{title}{Mechatronics- Electromechanics and contromechanics
  (Book)}.
\newblock \bibinfo{year}{1993}.
\bibitem[{mat(2017)}]{matlab}
\bibinfo{title}{{Mathworks Documentation for MATLAB R2017a}}.
\newblock
  \bibinfo{howpublished}{\url{https://www.mathworks.com/help/control/ref/pole.html}};
  \bibinfo{year}{2017}.
\bibitem[{Singh and Vadali(1993)}]{Singh93_ASME2}
\bibinfo{author}{Singh\xfnm[ T.]}, \bibinfo{author}{Vadali\xfnm[ S.R.]}.
\newblock \bibinfo{title}{Robust time delay control}.
\newblock \bibinfo{journal}{ASME Journal of Dynamic Systems, Measurement and
  Control}
  \bibinfo{year}{1993};\bibinfo{volume}{115}(\bibinfo{number}{2(A)}):\bibinfo{pages}{303--306}.
\bibitem[{Singh(2009)}]{singh2009optimal}
\bibinfo{author}{Singh\xfnm[ T.]}.
\newblock \bibinfo{title}{Optimal reference shaping for dynamical systems:
  theory and applications}.
\newblock \bibinfo{publisher}{CRC Press}; \bibinfo{year}{2009}.
\bibitem[{Nouh(2017)}]{nouh2017spatial}
\bibinfo{author}{Nouh\xfnm[ M.]}.
\newblock \bibinfo{title}{On the spatial sampling and beat effects in discrete
  wave profiles of lumped acoustic metamaterials}.
\newblock \bibinfo{journal}{The Journal of the Acoustical Society of America}
  \bibinfo{year}{2017};\bibinfo{volume}{141}(\bibinfo{number}{3}):\bibinfo{pages}{1514--1522}.
\bibitem[{Martin(1978)}]{Martin78}
\bibinfo{author}{Martin\xfnm[ G.]}.
\newblock \bibinfo{title}{On the control of flexible mechanical systems}.
\newblock Ph.D. thesis; Stanford University; \bibinfo{year}{1978}.
\bibitem[{Wie and Bryson(1981)}]{Wie1981}
\bibinfo{author}{Wie\xfnm[ B.]}, \bibinfo{author}{Bryson\xfnm[ J.]}.
\newblock \bibinfo{title}{{Modeling and control of flexible space structures}}.
\newblock In: \bibinfo{booktitle}{Dynamics and control of large flexiable
  aircrafts}. \bibinfo{year}{1981}, p. \bibinfo{pages}{153--174}.

\end{thebibliography}

\end{document}